\documentclass{aa}  
\usepackage{graphicx}
\usepackage{gensymb}
\usepackage{txfonts}
\usepackage{hyperref} 
\usepackage[table,xcdraw]{xcolor}

\usepackage{longtable}
\usepackage{pdflscape}
\usepackage{booktabs} 
\usepackage{dblfloatfix} 
\usepackage{afterpage}

\newcommand\kms{km\,s$^{-1}$}

\newcommand\percc{cm$^{-3}$}

\newcommand\micron{$\mu$m}

\newcommand\hii{H{\sc ii}}

\newcommand\ci{[C{\sc i}]}

\newcommand\CIIshort{[C{\sc ii}]}
\newcommand\CII{[C{\sc ii}]\,158\,${\rm \mu m}$}

\newcommand\OIshort{[O{\sc i}]}
\newcommand\OI{[O{\sc i}]\,63\,${\mathrm{\mu m} }$}
\newcommand\OIIIlong{[O{\sc iii}]\,88\,${\mathrm{\mu m} }$}
\newcommand\OIother{[O{\sc i}]\,145\,${\mathrm{\mu m} }$}

\newcommand\NIIlong{[N{\sc ii}]\,205\,${\mathrm{\mu m} }$}

\newcommand\NII{[N{\sc ii}]\,205\,${\mathrm{\mu m} }$}

\begin{document}
   \title{Uncovering the absorbed atomic Universe with the [OI]63$\mu$m line}

\author{Carlos De Breuck\inst{1}
\and Kevin~C. Harrington\inst{2,3,4,5}
\and Wout Hermans\inst{6}
\and Luke Maud\inst{1}
\and Aniket Bhagwat\inst{7}
\and Ilse De Looze\inst{6}
\and Bo Peng\inst{7}
\and Amit Vishwas\inst{8}
\and Benedetta Casavecchia\inst{7}
\and Andreas Lundgren\inst{9}
          }

\institute{European Southern Observatory, Karl Schwarzschild Stra\ss e 2, 85748 Garching, Germany;  \email{cdebreuc@eso.org}\label{inst1}
\and
Joint ALMA Observatory, Alonso de C{\'o}rdova 3107, Vitacura, Casilla 19001, Santiago de Chile, Chile\label{inst2},
\and              
National Astronomical Observatory of Japan, Los Abedules 3085 Oficina 701, Vitacura 763 0414, Santiago, Chile\label{inst3}
\and
European Southern Observatory, Alonso de C{\'o}rdova 3107, Vitacura, Casilla 19001, Santiago de Chile, Chile\label{inst4}
\and
Instituto de Estudios Astrofísicos, Facultad de Ingeniería y 455 Ciencias, Universidad Diego Portales, Av. Ejército Libertador 441, Santiago, Chile\label{inst5}
\and
Sterrenkundig Observatorium, Ghent University, Krijgslaan 281—S9, B-9000 Gent, Belgium
\and
Max Planck Institute f\"ur Astrophysik,  Karl-Schwarzschild-Strasse 1, 85748 Garching, Germany
\and
Cornell Center for Astrophysics and Planetary Sciences, Cornell University, Ithaca, NY, 14853, USA\label{inst6}
\and
Aix Marseille Univ, CNRS, CNES, LAM, Marseille, France}

   \date{Received 2025 December 19; accepted 2026 February 24}
 
  \abstract
  {We report the discovery of strongly absorbed \OI{} in a sample of 12 dusty star-forming galaxies (DSFGs) at 4.2$<$$z$$<$5.8. This is the first systematic survey of the \OI{} fine-structure line at $z>4$, targeting a sub-sample of gravitationally lensed DSFGs selected from the South Pole Telescope survey. Using ALMA Bands~9 and~10, we obtain spatially and spectrally resolved observations that probe the interstellar medium on sub-kiloparsec scales. Despite reaching sensitivities one to two orders of magnitude deeper than most previous studies, we detect \OI{} in emission in only two sources at low significance, with the remaining galaxies yielding stringent non-detections over the full velocity range covered by robust detections of other far-infrared lines, including \CII{} and \NII{}.
  We identify several compact (0.05-0.2$\arcsec$) regions having \OI{} absorption against the far-infrared dust continuum, some of which are possibly reaching below rest-frame Cosmic Microwave Background (CMB) radiation level. This suggests the presence of low-excitation temperature (Tex $\leq$ $T_{\rm CMB}$($z$)), low-density gas along those lines of sight. We also detect narrow, spatially localised \OI{} emission "escape channels" preferentially detected in regions with weak or absent dust continuum emission.  We also predict that similar absorption effects may appear in the \CII\ line, particularly when concentrating on the regions with the densest foreground material along the line of sight. 
  The \OI{} line appears to be strongly affected by the influence of extended star forming regions, with a mix of compact, high optical depth \OI{} emitting regions and sub-thermally excited, oxygen-rich  molecular clouds dispersed throughout high-redshift starbursts that are capable of absorbing the ground-state line emission. Combined with a comparison to cosmological radiation hydrodynamical simulations, this supports the interpretation that regions with higher gas and dust column densities may lead to weakening an intrinsically strong \OI{} line emission. We argue that the high \OI{} optical depth is the dominant effect causing the strong absorption, limiting the diagnostic power of this line to trace regions of massive star formation in high-redshift DSFGs. 
}

               
\maketitle

\section{Introduction}

Far-infrared (FIR) fine-structure lines (FSL) are among the major coolants of the interstellar medium (ISM), typically emitting at a few percent of the total infrared luminosity in galaxies \citep[e.g.][]{Crawford1985,Stacey1991,Fischer1999,madden2013,cormier2015,Wolfire2022}. These lines are also excellent tracers of the physical conditions in the various phases of the ISM \citep{Abel2005,Nagao2011,Decarli2025,Peng2025c}. The \CII{} line is one of the primary coolants and is not significantly affected by dust attenuation due to its relatively long wavelength. For almost four decades, it has been a workhorse for extragalactic studies near and far and used as a tracer of star formation activity \citep[e.g.][]{Brauher2008,stacey2010,DeLooze2014,capak_2015}. Due to its low excitation potential (91\,K above ground), the \CIIshort{} line has also been routinely used to trace extended gas in galaxy haloes \citep[e.g.][]{HerreraCamus2021,Lambert2023} and to the cold gas reservoir of galaxies \citep{zanella_2018,Madden2020}. 
Other FSL such as \OIIIlong{} and 52\,$\mu$m, \NIIlong{} and 122\,\micron{} trace the \hii{} regions, while \CIIshort{} originates from both the neutral and ionised gas phases, given that the ionisation potential of carbon (11.26 eV) is lower than that of hydrogen (13.6 eV). On the other hand, the \OIshort{} emission is thought to primarily originate from dense photo-dissociation regions \citep[PDR;][]{Tielens1985,Malhotra2001}, and are frequently used in the local Universe to constrain the radiation field intensity ($G_{\text{0}}$) in combination with the \CIIshort{} line and total infrared (TIR) dust continuum emission. 

The \OIshort\,63\,\micron{} and 145\,\micron{} lines probe the ISM in the high density regime due to the high critical densities for collisions with H atoms of 4.7 or 7.8$\times$10$^{5}$ cm$^{-3}$ for the \OI\ line and 0.95 or 2.0$\times$10$^{5}$ cm$^{-3}$ for the \OIshort\,145\,$\mu$m line, reporting the values by \citet{fernandezontiveros2016} and \citet{Goldsmith2019}, respectively. Because the \OI\ line emission is due to the ground-state transition (${}^3P_1 \rightarrow {}^3P_2$) of neutral oxygen with a high Einstein A coefficient, and the neutral gas is often arranged in dense clouds that have high column density, the \OI{} line is often optically thick (i.e., not all photons are able to escape from the cloud), and prone to self-absorption effects (due to the absorption of 63\,$\mu$m photons by material along the line of sight). The \OIshort\,145\,$\mu$m line transition (${ }^3P_0 \rightarrow {}^3P_1$) has a de-excitation coefficient that is $\sim$50$\times$ smaller than the 63\,$\mu$m line \citep{Lique2018}, and is hence optically thin in almost all astrophysical environments \citep[e.g.][]{Goldsmith2019}. 
As a result, the ratio of both lines, i.e., \OI\ / \OIshort\,145$\mu$m is often affected by optical depth due to self-absorption effects; only in cases where the 63\,$\mu$m line is optically thin can the line ratio be used to determine the gas density in the high density regime.
Galaxy zoom-in simulations also predict the \OI\ and 145\,$\mu$m lines to play an increasingly important role as cooling lines in the neutral gas and should be readily detectable in high-redshift galaxies \citep{Pallottini2022}. 
Being prone to optical depth and self-absorption effects, it is reasonable to consider that the apparent \OI\ line emission could be significantly diminished compared to theoretical expectations in some of the most compact dusty high-z galaxies. 

The \OIshort{} 145$\mu$m line has been routinely detected in local galaxy samples, ranging from low-metallicity dwarfs \citep{cormier2015} over normal star-forming galaxies \citep{Herrera2018} to starbursts \citep{Farrah2013} and AGN-dominated galaxies \citep{Spinoglio2015,fernandezontiveros2016}. In recent years, the \OIshort\,145$\mu$m has also been detected in some high-redshift quasars and starbursts \citep{debreuck2019,Yang2019,Lee2021,Meyer2022,Litke2023,Fudamoto2025}. The detections of the \OI\ line have remained limited mostly to $z \sim 2$ galaxies detected with {\it Herschel} \citep{Coppin2012,Brisbin2015,Wardlow2017,Zhang2018,Wagg2020}. Despite having the capability to observe \OI\ at $z>4$ since 2012, the first ALMA detections were only recently reported in J2054-0004 at $z$=6.040 \citep{Ishii2025} and W2246-0526 at $z$=4.601 \citep{FernandezAranda2024}. In the latter source, the \OI\ line luminosity rivals that of the \CIIshort\ line emission; however, an earlier bright \OI\ detection \citep{Rybak2020} was later refuted \citep{Rybak2023}, showing that the undetected \OI\ line was consistent with being at least 5 times fainter than \CIIshort. Also in SPT0346$-$52 at $z$=5.7, the \OI\ was not detected\footnote{At this redshift, the \OI\ falls at frequency with a particularly low atmospheric transmission, limiting the sensitivity.} with a luminosity $<$0.6$\times$ that of \CII\ \citep{Litke2022}. However, we must consider two biases potentially affecting the current picture in the literature: first,  publication bias leads to non-detections being under-represented in papers, and second, the \OI\ line can only be observed in the high frequency bands 9 or 10 at $4<z<6.88$, which were perceived to be more difficult and hence under-utilised compared to the lower frequency Bands 6, 7 and 8 where the \CII\ and \OIother\ lines are  observable at $z>2.8$ and $z>3.1$, respectively. The view of \OI{} at high-$z$ is still highly incomplete, and appears so far different from our knowledge in the local universe. Clearly, a complete and statistical sample of \OI\ observations at $z>4$ is required to determine the origin of this difference with low redshift samples.

In this paper, we report the results from the first survey of \OI\ at $z>4$ from gravitationally-lensed, dusty star-forming galaxies (DSFGs) selected from the South Pole Telescope \citep[SPT;][]{Reuter2020}. This sample is unique in both the 10-band photometric coverage of the dust Spectral Energy Distribution (SED) from 3\,mm to 70\,$\mu$m, its complete redshift coverage and rich ancillary data covering many FSLs such as \ci\ \citep{Bothwell2017}, \CII\ \citep{gullberg2015}, and \NII\ \citep{Cunningham2020}, as well as multiple CO transitions \citep{weiss_2013,strandet2016,Aravena2016,Reuter2020}. From this sample, we selected all 20 targets where the \OI\ line falls within the atmospheric windows accessible to ALMA.

This paper is structured as follows. In Sect.~2 we describe both the \OI\ and archival observations. Sect.~3 presents our results, which are discussed in Sect.~4. Sect.~5 contains our conclusions and summary. Throughout this paper, we adopt the following cosmology: $H_0 = 67.8\, \mathrm{km\,s^{-1}\,Mpc^{-1}}$, $\Omega_m = 0.308$, and $\Omega_\Lambda = 0.692$ \citep{Planck2016}.

\section{Observations}
\subsection{Band 10 and 9 \texorpdfstring{\OIshort{}}{[OI]} observations}\label{section:Band10observations}

The observations were conducted between 2024 October and 2025 April as part of project ID 2024.1.01465.S (PI: Harrington). We present data using the ALMA Band 10 receiver \citep{Uzawa2013} of ten sources at 4$<$$z$$<$5 and the Band 9 receiver \citep{Baryshev2015} of two sources at 5.6$<$$z$$<$6.9. As part of the project, Band 10 data were also taken of two additional unlensed sources from our sample and are reported elsewhere \citep[][Vishwas et al, in prep.]{Harrington2025}; two DSFGs (SPT0552-42 and SPT0544-40) were missed within the $\sim$13\arcsec ALMA primary beam due to the large positional uncertainty of the known SPT coordinates. Four sources in our project sample were not observed due to scheduling constraints or the absence of suitable calibrators. The sample of 12 lensed DSFGs presented here therefore represents an unbiased sub-sample.

Most observations were taken in configurations C43-2 and C43-3 with maximum baselines up to 500\,m, but the April data were taken in C43-5 with maximum baselines up to 1400\,m. Given that the typical spatial extent of the targets is $\sim$ 1--2\arcsec{}, they are smaller than the Maximum Recoverable Size (MRS), except for SPT2146-55, where some emission may have been resolved out ($\theta_{\rm MRS}$ $\sim$ 0\farcs{}8). 

The Band 9 and 10 ALMA receivers are double sideband (DSB), and thus while configuring the spectral windows with a maximum bandwidth of 1.875\,GHz in each signal sideband, 90\,\degree{} Walsh switching correlator mode can be enabled to recover the corresponding image sideband (enabling additional spectral coverage on the opposite side of the local oscillator frequency\footnote{\url{https://almascience.eso.org/proposing/proposers-guide}}). Our defined spectral setup consisted of 4 contiguous spectral windows of 1.875\,GHz bandwidth each defined in the signal sideband using Frequency Division Mode (FDM) providing 7.5\,GHz bandwidth per sideband, and a total of 15\,GHz bandwidth when the signal and image sidebands are combined for imaging. We chose the tuning of the spectral windows to minimise the impact of telluric absorption features in both the signal and image bands of the observing setup, as illustrated by the blue and grey curves in the top panels of Figs.~\ref{figure:SPT0106OI-overview} to \ref{figure:SPT2311OI-overview}, respectively. Despite the large bandwidth, calibration is often difficult for high frequency data, and the band-to-band calibration scheme was employed by the observatory for most observations - see \S\,\ref{section:high-frequency} for further details.

We obtained calibrated measurement sets provided by the ALMA pipeline, which was found to be reliable the for high-frequency data, including those using band-to-band (included since the 2024 pipeline version\footnote{\url{https://almascience.eso.org/processing/science-pipeline}}). In a few cases manual data reduction was provided by the European ALMA Regional Center due to lower signal-to-noise datasets requiring specific treatment to carefully flag data. Line emission is expected to span across spectral windows so we cannot use pipeline imaging products, which are produced on a per spectral window basis. We therefore used the \texttt{tclean} task in \texttt{CASA} \citep{casa_team_2022} with all spectral windows and an auto-multithresh \citep{kepley2020} cutoff. We used a range of spectral bin widths ranging from the native spectral resolution of a few \kms\ to 10, 20, 50 and 110\,\kms. In the remainder of this paper, we will use the 20 or 50\,\kms\ data, which was an optimal compromise between S/N and spectral resolution for narrow lines. For imaging the spectral line data, we used the {\tt robust=0.5} parameter, while for the continuum images, we maximised the S/N by using {\tt robust=2}. 

\subsection{High frequency considerations}\label{section:high-frequency}
One of the most important considerations for high frequency observations is that the phase rms is sufficiently low (stable atmosphere) as to minimise the loss of flux due to decoherence during phase correction, while a second is that close enough and bright enough phase calibrators are used for phase referencing. All but two sources were observed using the band-to-band calibration technique that observes a phase calibrator at a lower frequency than the high frequency tuning required for the science target. This method scales the high signal-to-noise solutions from the phase calibrator to the target source observed in the high frequency band during phase transfer, while also using a differential gain calibrator to align the band-to-band offset \citep[see.][]{asaki2020a,asaki2020b,asaki2023,maud2020,maud2022,maud2023}. The technique allows phase transfer using nearby phase calibrators for observations that would have calibrators that are too distant or too faint for in-band calibration.

Table\,\ref{table:observations} lists the conditions and parameters of the observations. As can be calculated by inputting the phase rms into Equation 1 of \citet{maud2022}, under the assumption of point source detection, the visibility coherence can be used to estimate the final image coherence. The phase rms is established using the high-frequency data of the bandpass calibrator by selecting time samples equivalent to the phase-referencing cycle-time\footnote{The time taken from the start of a phase calibrator scan, slew to the target, the target scan, and the slew back to the phase calibrator ready to start the next scan.}, thus acting as a proxy for the phase rms likely remaining in the science target(s) after phase referencing - see \citet{maud2022,maud2023rms} and \S\ 3.14.3 of \citet{Hunter2023} for additional details. For our data, although most fields are (sometimes marginally) resolved, this still provides a useful estimate of decoherence. Of our sources, SPT2132-58 has the worst phase rms of 47\degree{}, which relates to a visibility coherence of $\sim$0.71, i.e. approximately 29\,\% of the flux is not well `focused' in the target location within the image. All other observations have notably smaller phase rms and thus typically a flux loss of $\sim$10-15\,\% can be expected.  
When considering image degradation caused by high separation angles to the phase calibrator, previous high frequency studies focussed on longer-baseline observations \citep{maud2020,maud2023}. These authors find that the degradation worsens with larger separation angles as a function of increasing frequency and longer baseline lengths. Despite investigations lacking at Band 10 for short baselines, it should be reasonable to adopt a coherence degradation of $-$0.022 per degree separation to the phase calibrator as stated for `short' baselines under 5000$k\lambda$. As our observations typically have calibrators $<$4\degree, a general degradation of $<$9\,\% can be expected. For SPT2311-54, the phase calibrator separation angle to the target is $\sim$7\degree{} as this was a `shared' observation with another source where one common phase calibrator was used. In this case SPT2311-54 could have an additional degradation of $\sim$15\,\%. 
Considering most our observations use band-to-band, the investigations by \citet{maud2020} also stated there could be a degradation due to an underlying phase rms residual of the differential gain calibrator after frequency scaled phase correction. For a residual phase rms $<$20\degree{} (in the differential gain calibrator phases) the degradation amounts to $<$5\,\%. This is reasonable for our observations given the low to high frequency scans on the differential gain calibrator have a faster cycle time than the phase calibrator to the target, for which the estimated phase rms values indicated in Table~\ref{table:observations} are already low. Finally, we note that absolute astrometric accuracy also depends on the phase transfer and to some extent the decoherence. Typically, we do not expect the absolute position uncertainties to exceed 10-20\,\% of $\theta_{synth}$ - see \S\ 10.5.2 of the ALMA Technical Handbook \citet{cortes2025}.

Summarised, a combined flux loss of up to 10--30\% can be expected due to phase calibration, which is similar to the absolute flux calibration accuracy of $\sim$20\%. Appendix~\ref{section:flux-calibration} compares our Band 9 and 10 data with published {\it Herschel} data, yielding consistent photometry in half of our sources, where the lower ALMA fluxes in the other half may be due to the inclusion of companion sources in the much larger {\it Herschel} beam.

\begin{table}
\caption{ALMA Band 9 and 10 observations} 
\centering    
\resizebox{\columnwidth}{!}{
\begin{tabular}{c c c c c c c c}       
\hline\hline                
Source & $z$ & $\nu_{\rm obs}$ & $t_{\rm int}$ & $\theta_{\rm synth}$ & PWV & phase  & phase cal\\   
 &  & GHz & s & $\arcsec \times \arcsec$ & mm & rms  & separation \\   
\hline                       
SPT0106-64 & 4.910  & 811.9 & 1022 & 0.27$\times$0.15 & 0.32 & 14\degree &0.5\degree\\     
SPT0136-63 & 4.299  & 886.6 & 2586 & 0.19$\times$0.13 & 0.52 & 31\degree &4.0\degree\\
SPT0155-62 & 4.349  & 880.1 & 1084 & 0.14$\times$0.12 & 0.38 & 18\degree &5.9\degree\\
SPT0243-49 & 5.702  & 697.6 &  897 & 0.13$\times$0.10 & 0.37 & 19\degree &5.5\degree\\
SPT0418-47 & 4.225  & 899.4 & 1080 & 0.17$\times$0.13 & 0.45 & 16\degree &5.7\degree\\
SPT0441-46 & 4.4803 & 859.4 &  929 & 0.10$\times$0.09 & 0.65 & 30\degree &2.6\degree\\
SPT0459-58 & 4.856  & 803.4 & 2290 & 0.19$\times$0.15 & 0.61 & 31\degree &3.2\degree\\
SPT0459-59 & 4.799  & 811.3 & 1856 & 0.19$\times$0.15 & 0.62 & 36\degree &3.2\degree\\
SPT2132-58 & 4.768  & 813.7 & 2416 & 0.05$\times$0.04 & 0.64 & 47\degree &3.5\degree\\
SPT2146-55 & 4.567  & 846.4 & 1176 & 0.04$\times$0.03 & 0.39 & 32\degree &3.3\degree\\
SPT2311-54 & 4.2795 & 888.6 & 2167 & 0.16$\times$0.10 & 0.41 & 40\degree &7.0\degree\\
SPT2351-57 & 5.811  & 687.7 & 2150 & 0.30$\times$0.16 & 0.52 & 13\degree &4.3\degree\\
\hline                                   
\end{tabular}
}

\tablefoot{Redshifts taken from \citet{Reuter2020}. Synthesized beam assuming {\tt robust=0.5}. Precipitable Water Vapour (PWV) and phase rms is the highest value, rounded to the nearest degree in case of multiple executions, as extracted from the bandpass calibrator high-frequency data and sampled over the phase-referencing cycle-time}.
\label{table:observations}    
\end{table}
\subsection{Supporting archival data}
Seven of the 12 sources with \OI\ data already have published \CII\ data from the Atacama Pathfinder Experiment (APEX) \citep{gullberg2015}. The remaining five sources were observed between 2021 October and November with the ALMA 7m array (ID 2021.1.00857.S, PI: De Breuck) using Band 7 \citep{Mahieu2012}. Each source was observed for 33 minutes using a tuning covering the \CII\ line. The data were reduced by the ALMA pipeline, and re-imaged as described in §~\ref{section:Band10observations}.
The synthesized beam of the 7m array at these frequencies is 2\farcs{}1--2\farcs{}6, meaning that all our sources are unresolved, providing a total line flux similar to the APEX spectra.

Nine of the sources also have \NII\ data observed with the ALMA 7m array \citep{Cunningham2020}. For consistency, we obtained calibrated measurements sets from the Additional Representative Images for Legacy \citep[ARI-L;][]{Massardi2021}, re-imaged these data using the same procedures as in §~\ref{section:Band10observations}, and list the newly derived flux densities in Table~\ref{table:measurements}. To compare the continuum morphology, we downloaded archival Band 7 data, part of which were previously published by \citet{Spilker2016}.

\section{Results}
\subsection{Continuum results}
The central panels of Figs.\,\ref{figure:SPT0106OI-overview} to \ref{figure:SPT2311OI-overview} show the {\it restframe} 63\,\micron{} continuum emission\footnote{Exact observed frequencies at listed in column 3 of Table~\ref{table:observations}.} with archival {\it observer frame} 870\,$\mu$m continuum emission overlaid as white contours. As the synthesized beam sizes are very similar, these overlays allow a direct comparison, showing very consistent overall morphologies between both bands, providing confidence in the quality of the high-frequency data. Looking into the finer details, we do notice some variations in the locations of the dust peaks (e.g. SPT0418-47). However, as the Band 10 data covers $\sim$3$\times$ higher frequencies where the dust emission may become optically thick, a detailed radiative transfer treatment is required; we defer such an investigation to a future publication.

\subsection{\texorpdfstring{\OIshort{}}{[OI]} results}

\begin{figure*}
\centering
\includegraphics[width=15cm]{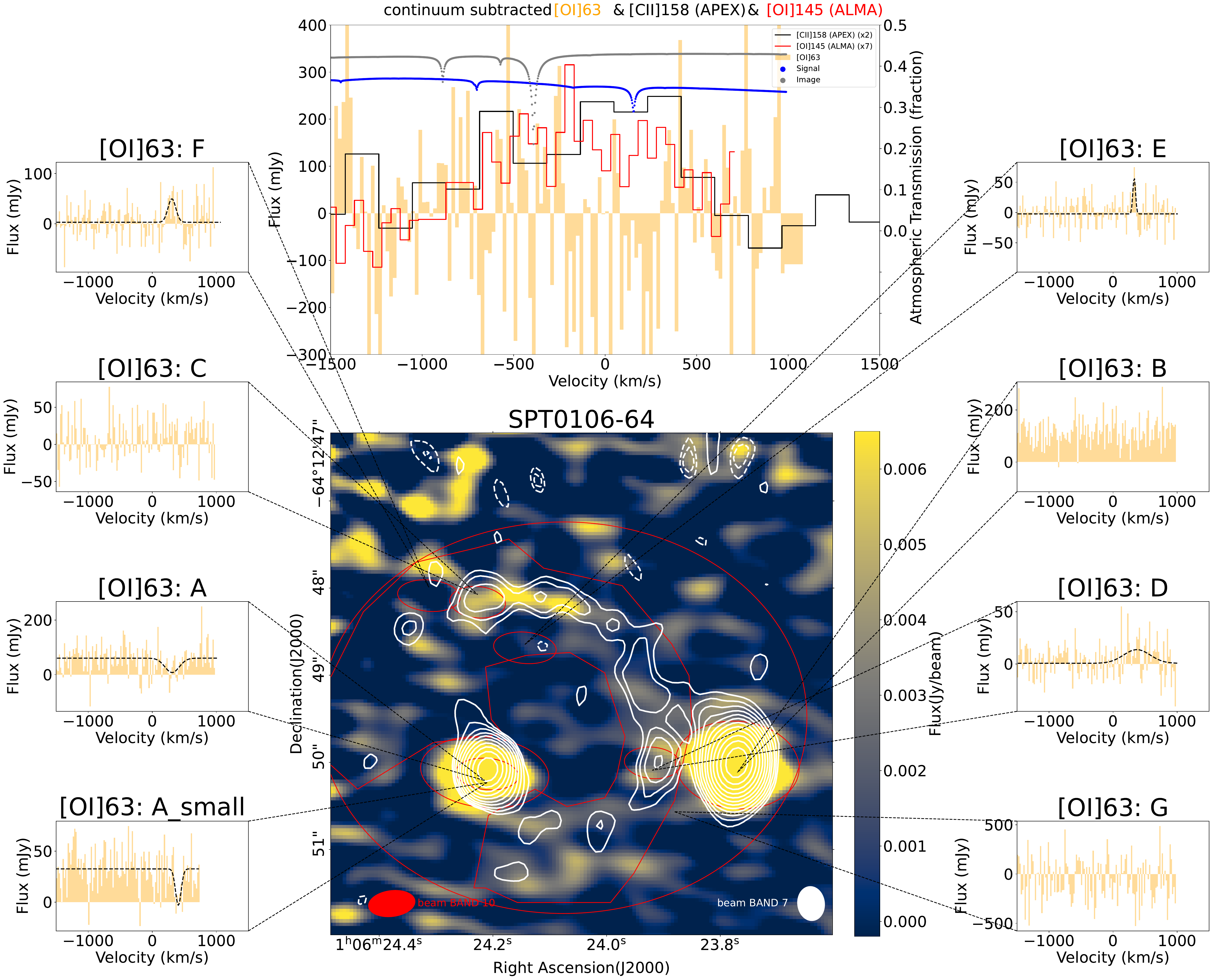}\\
\vspace{0.5cm}
\includegraphics[width=15cm]{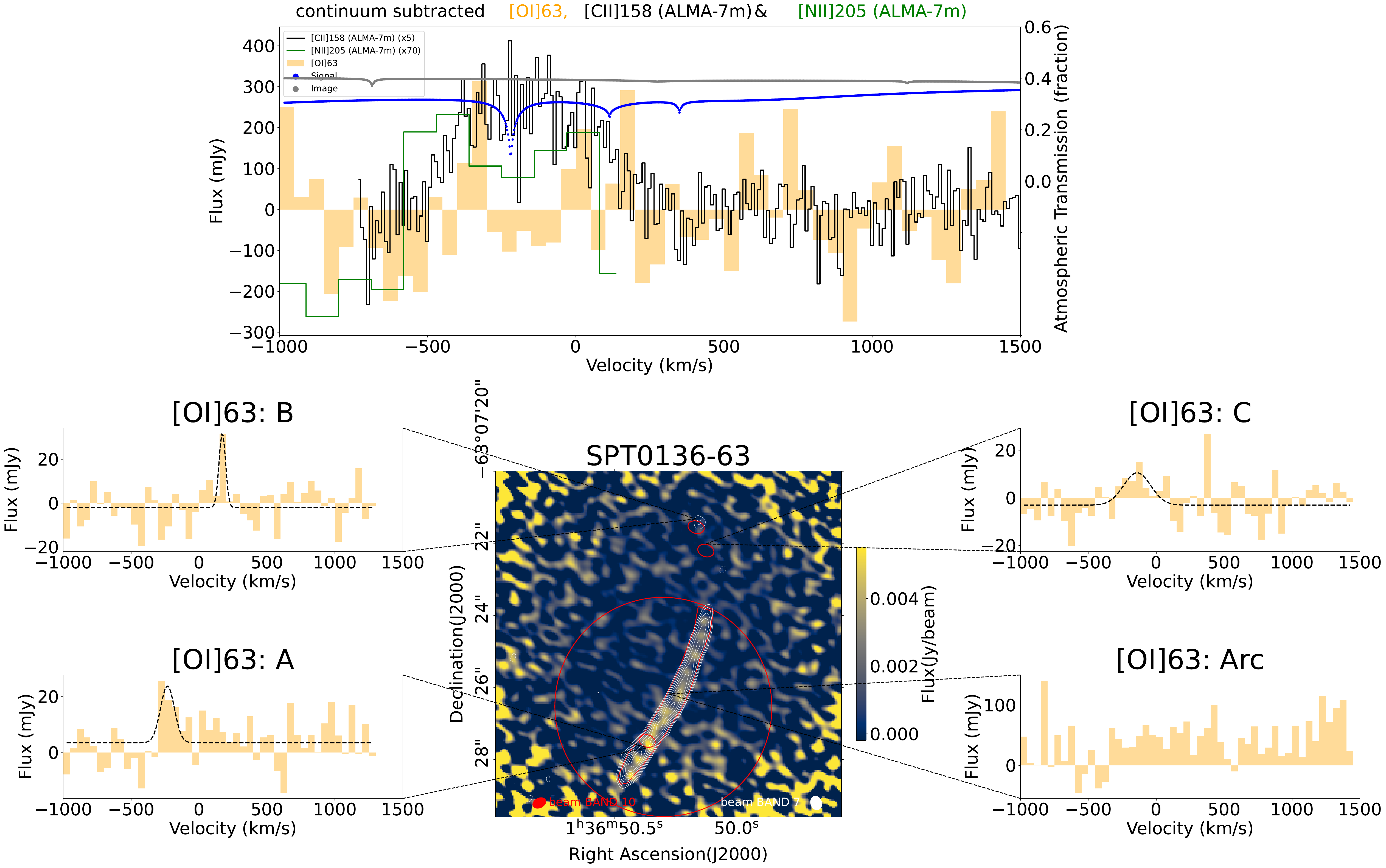}
      \caption{Overview of the results for SPT0106-64 and SPT0136-63 The central panel shows the Band 10 or 9 line+continuum image (at central frequency listed in Table\,\ref{table:observations}) in colour with the archival Band 7 continuum in white contours for reference. The beam sizes for both images are shown as red and white filled ellipses. The apertures used in Table\,\ref{table:measurements} are marked in red, and the extracted line+continuum spectra are shown in the left and right side panels. The top panel shows the continuum-subtracted \OI\ spectrum in yellow with the appropriate atmospheric transmission in both the signal and image band overplotted in blue and grey, respectively. The \CII\ spectra from APEX or the ALMA 7m-array are overplotted in black, and where available, the \NII\ spectrum is overplotted in green, with the scaling factors marked in the legend.}
         \label{figure:SPT0106OI-overview}
\end{figure*}

\begin{figure*}
\centering
\includegraphics[width=16cm]{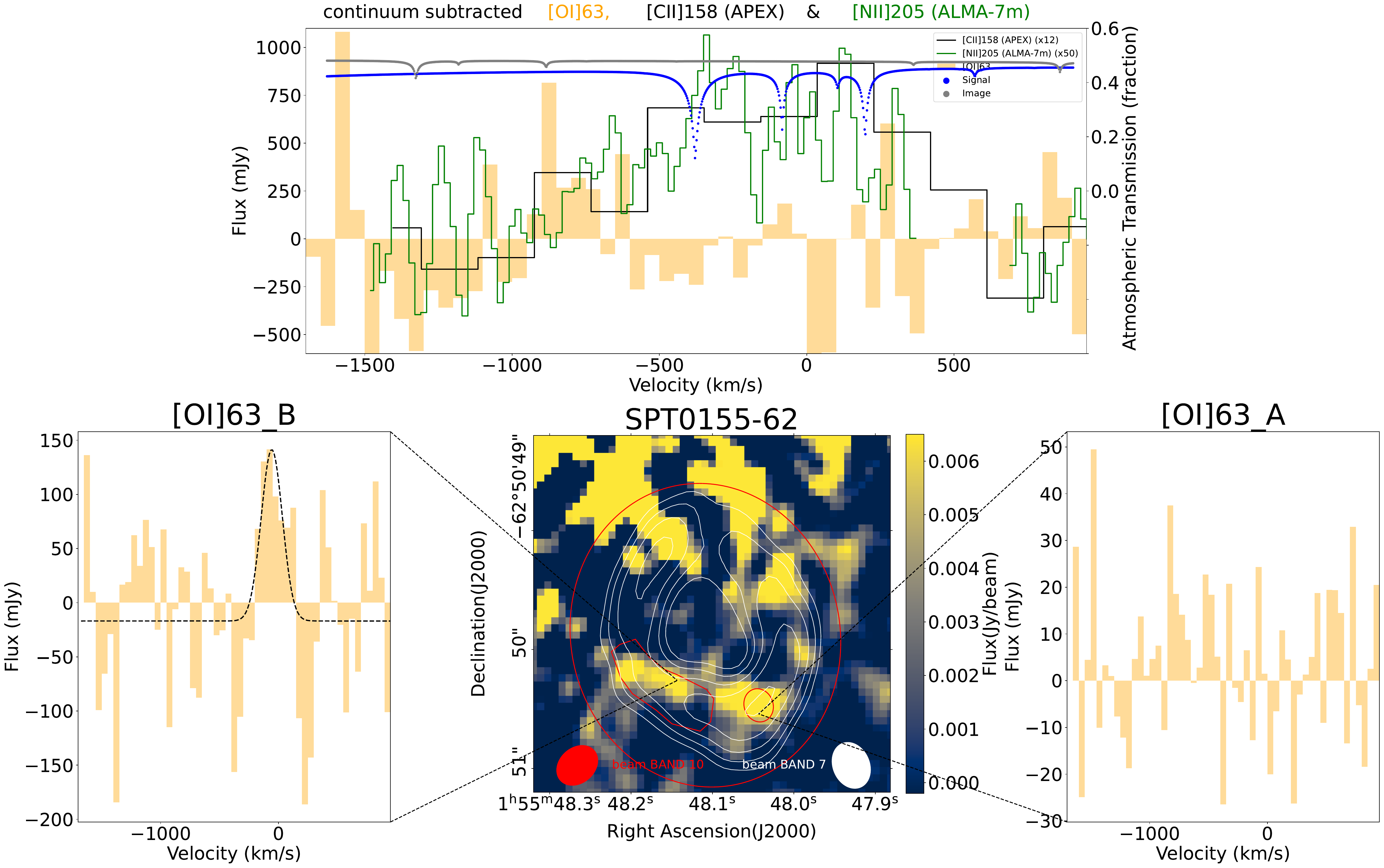}\\
\vspace{0.5cm}
\includegraphics[width=16cm]{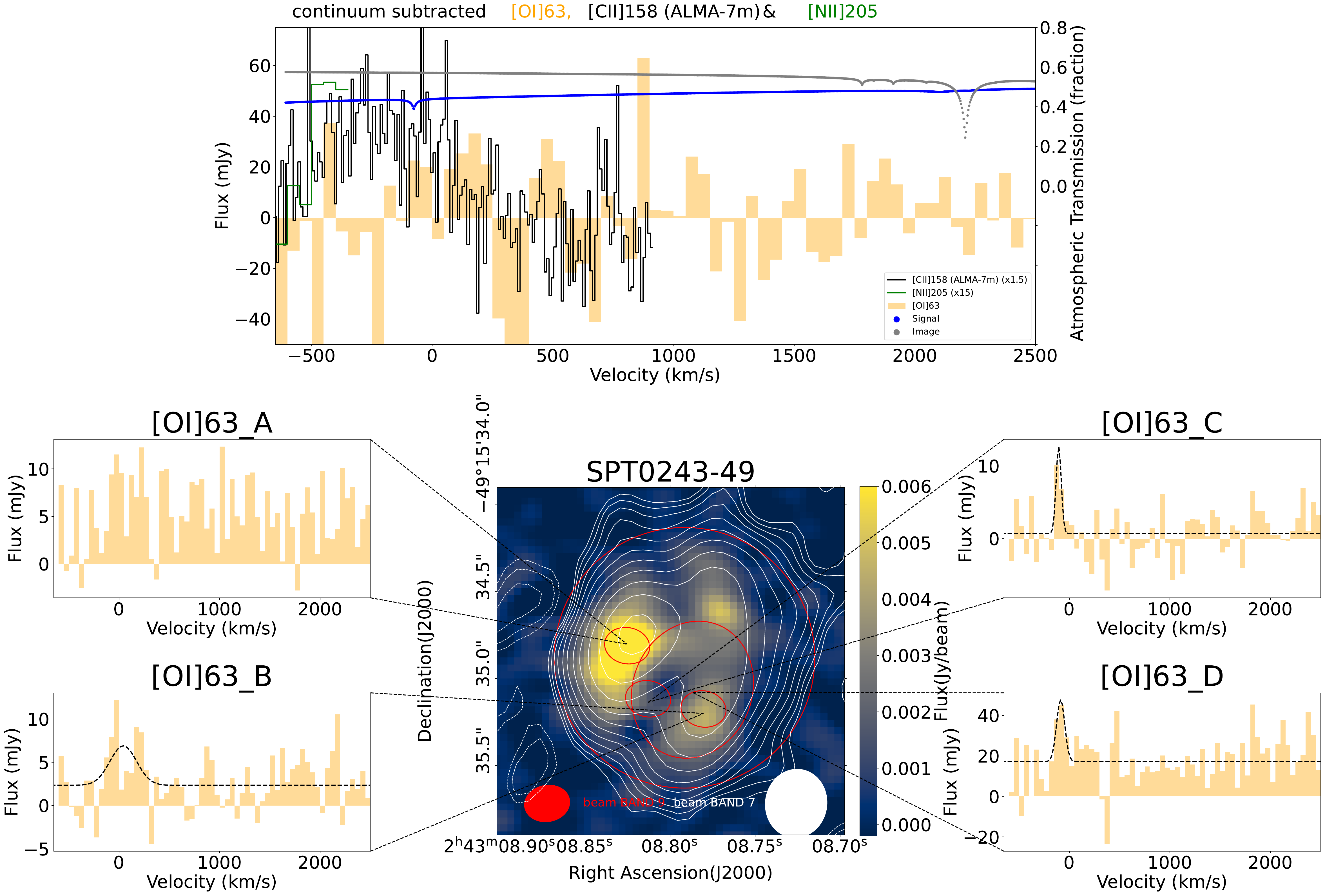}
      \caption{Overview of SPT0155-62 and SPT0243-49 as in  Fig.~\ref{figure:SPT0106OI-overview}.}
      \label{figure:SPT0155OI-overview}
\end{figure*}

\begin{figure*}
\centering
\includegraphics[width=15cm]{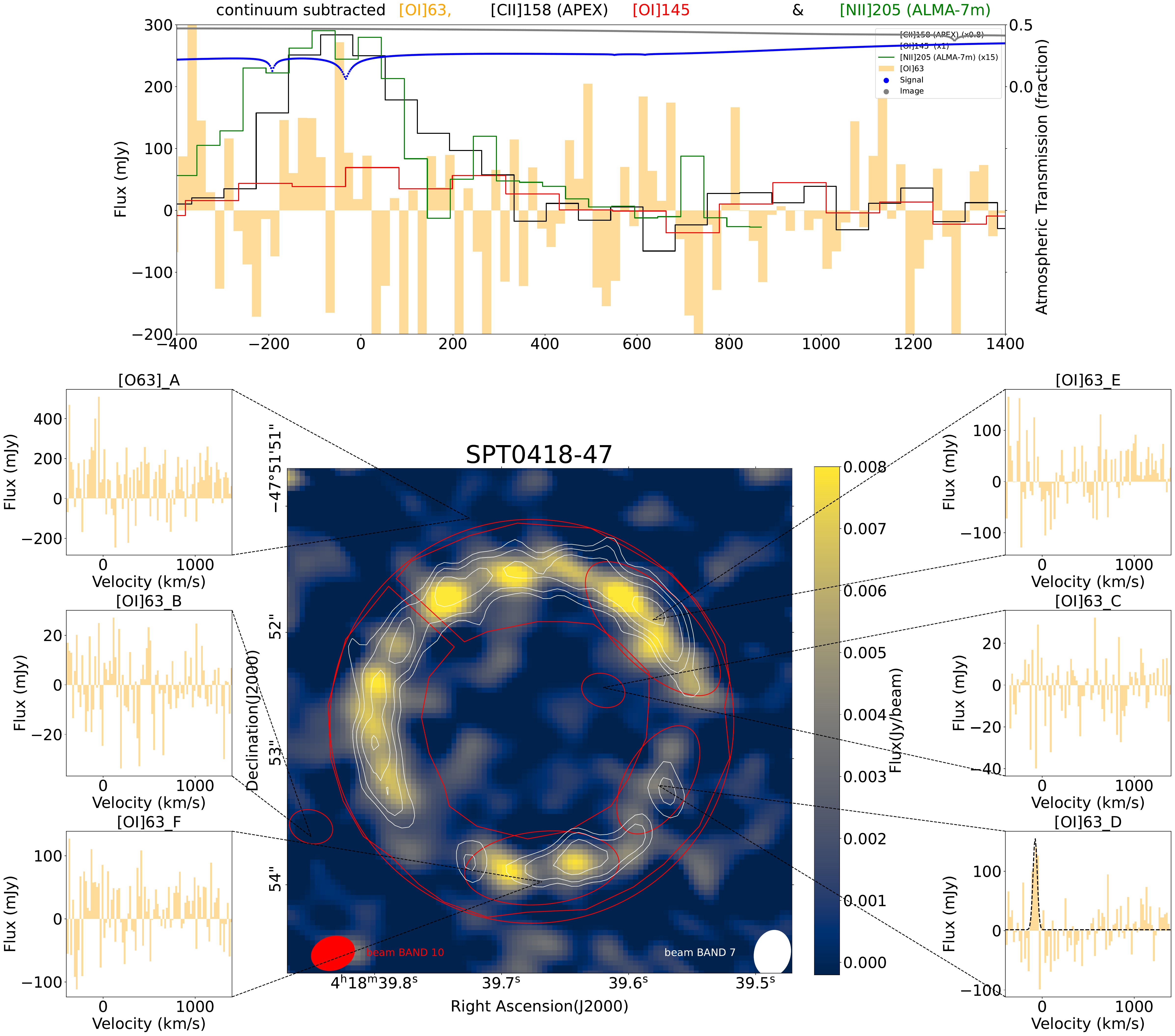}\\
\vspace{0.5cm}
\includegraphics[width=15cm]{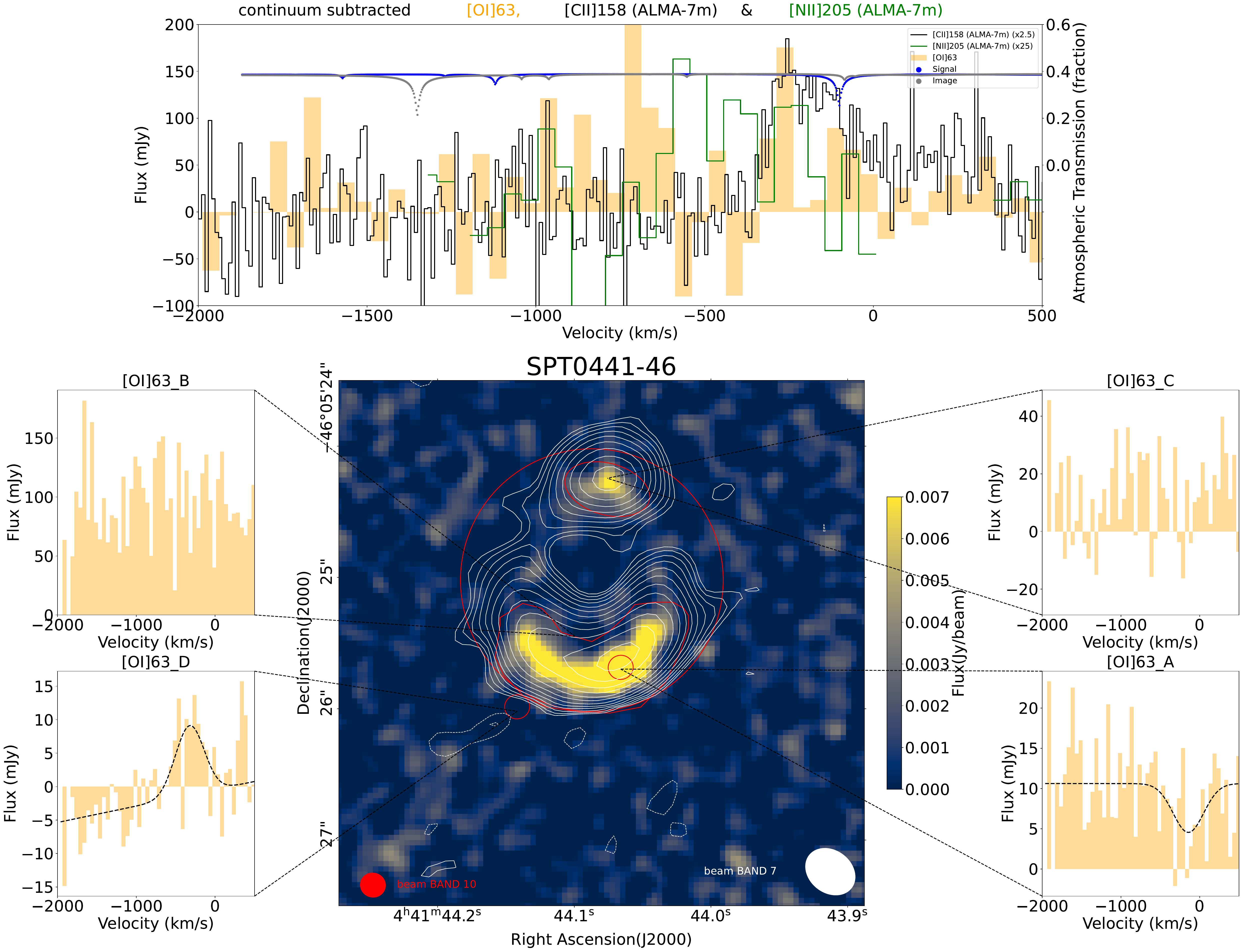}
      \caption{Overview of SPT0418-47 and SPT0441-46 as in  Fig.~\ref{figure:SPT0106OI-overview}.}
      \label{figure:SPT0418OI-overview}
\end{figure*}

\begin{figure*}
\centering
\includegraphics[width=15cm]{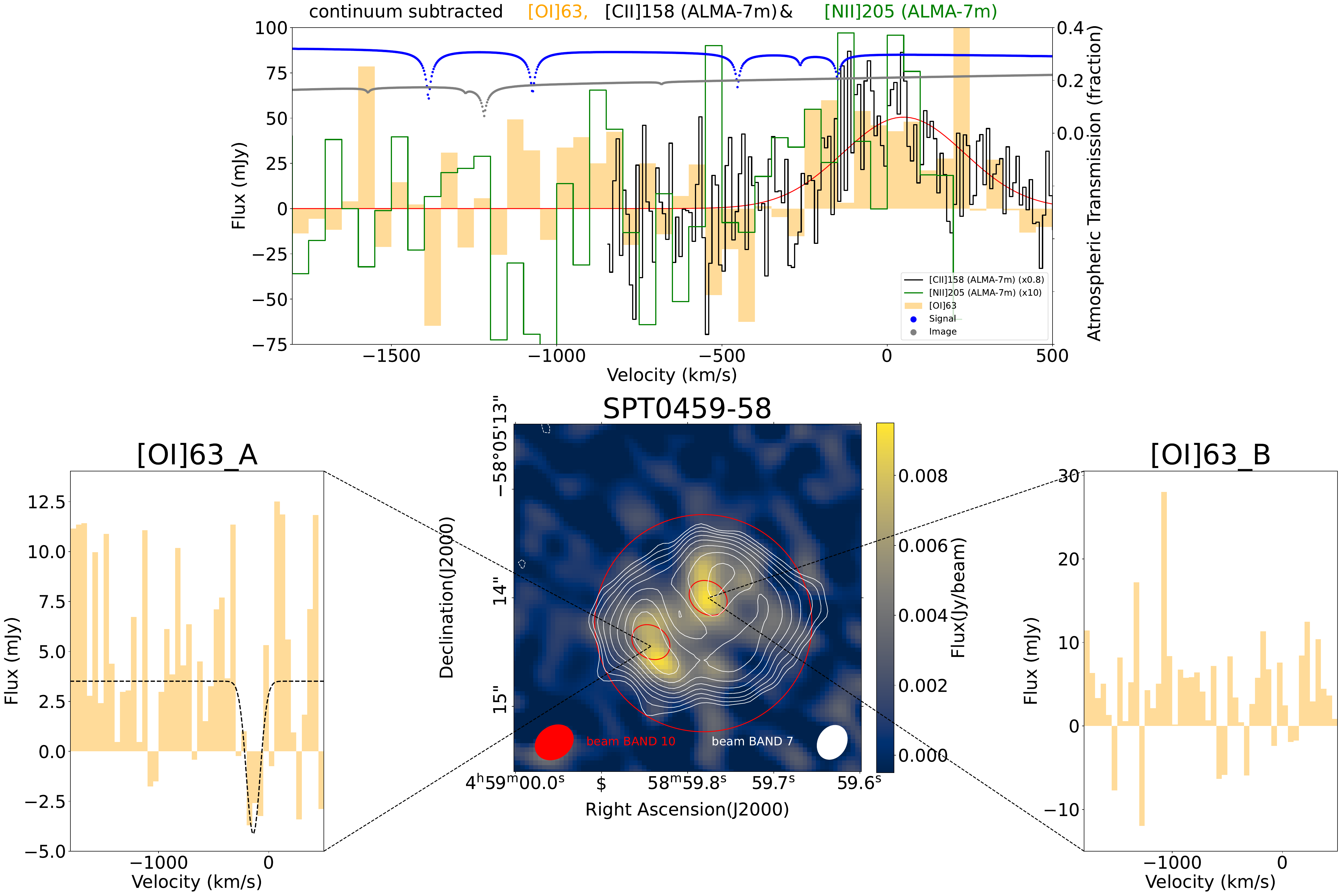}\\
\vspace{0.5cm}
\includegraphics[width=15cm]{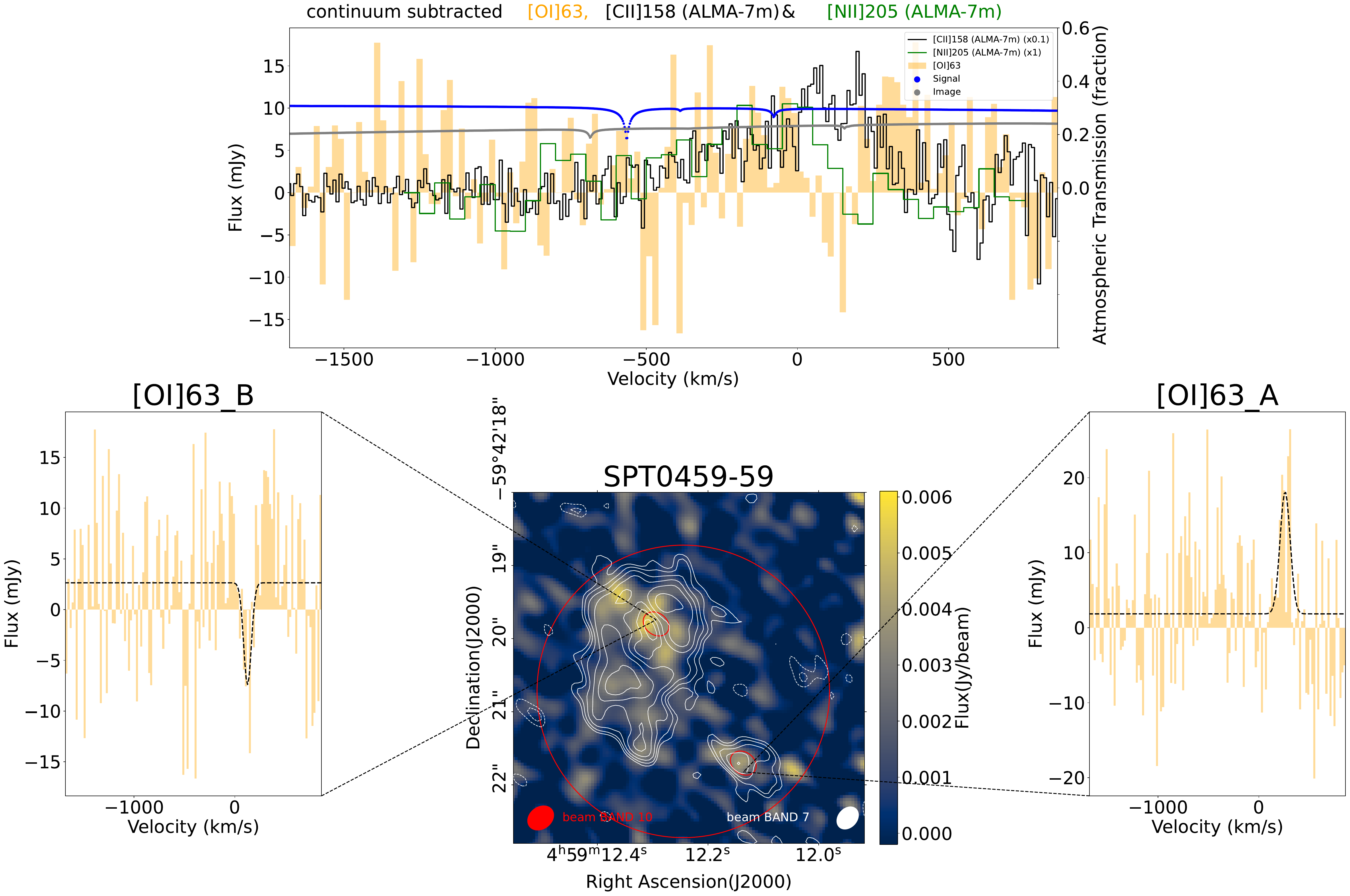}
      \caption{Overview of SPT0459-58 and SPT0459-59 as in  Fig.~\ref{figure:SPT0106OI-overview}.}
      \label{figure:SPT0459OI-overview}
\end{figure*}

\begin{figure*}
\centering
\includegraphics[width=16cm]{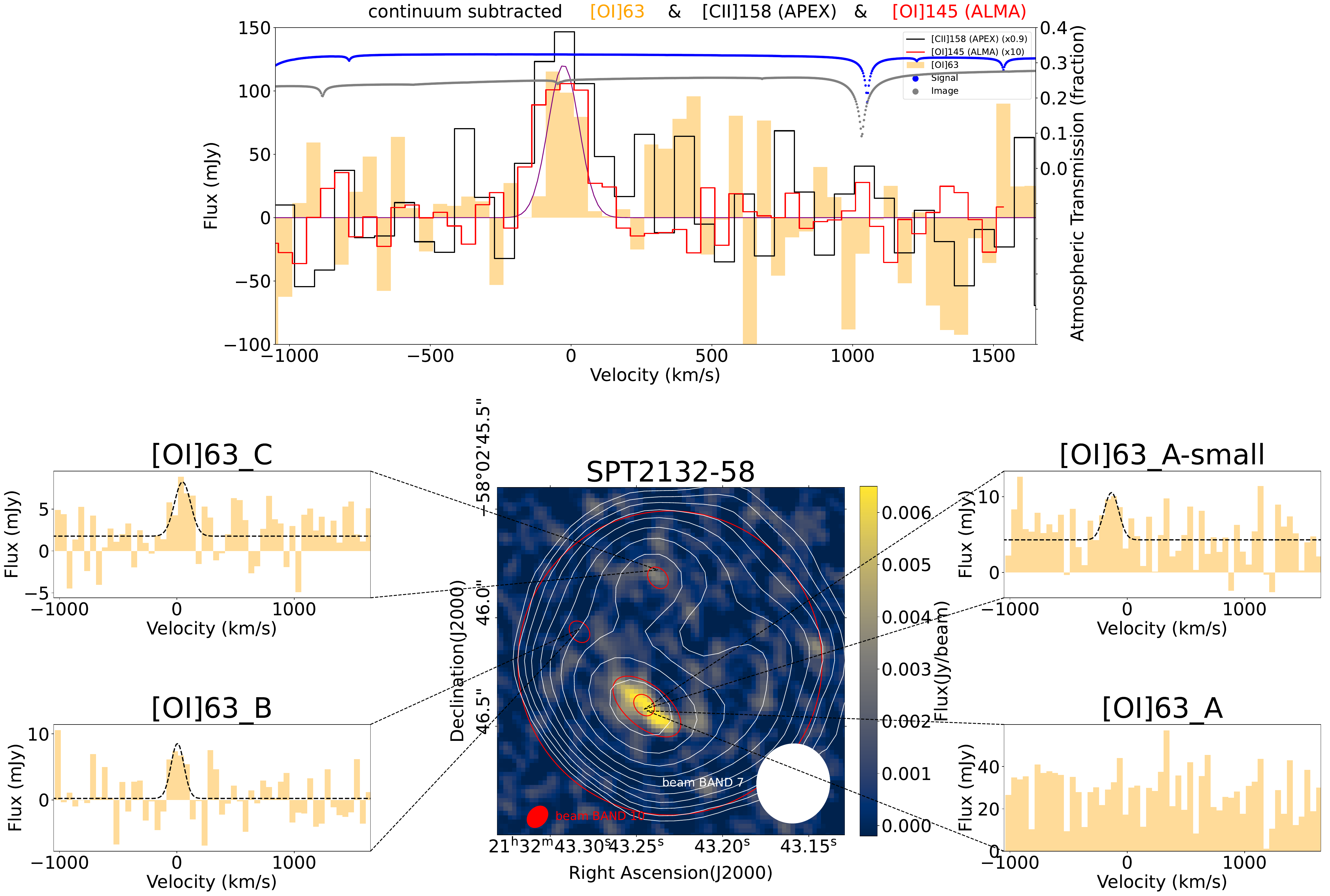}\\
\vspace{0.5cm}
\includegraphics[width=16cm]{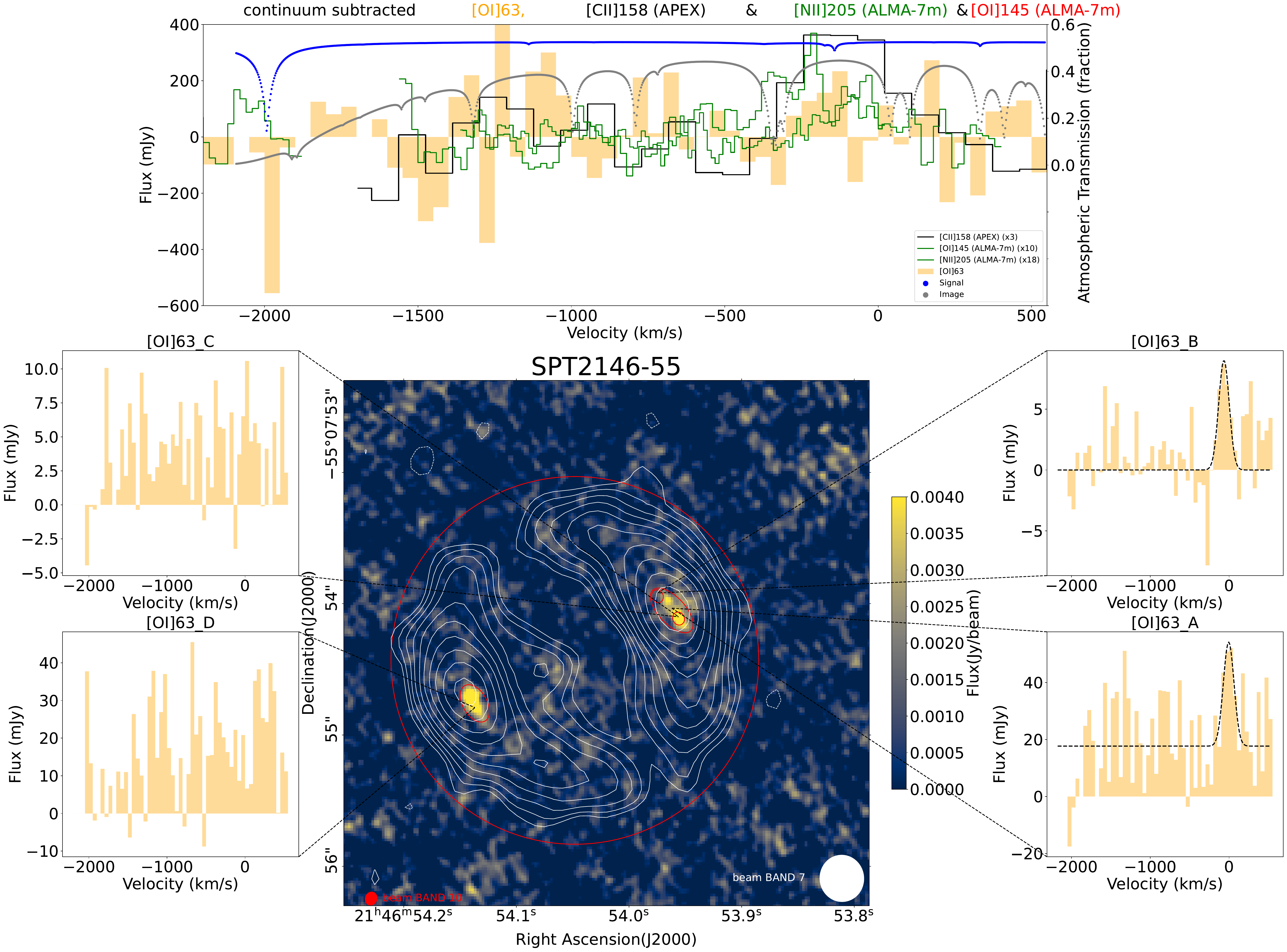}
      \caption{Overview of SPT2132-58 and SPT2146-55 as in  Fig.~\ref{figure:SPT0106OI-overview}.}
      \label{figure:SPT2132OI-overview}
\end{figure*}

\begin{figure*}
\centering
\includegraphics[width=16cm]{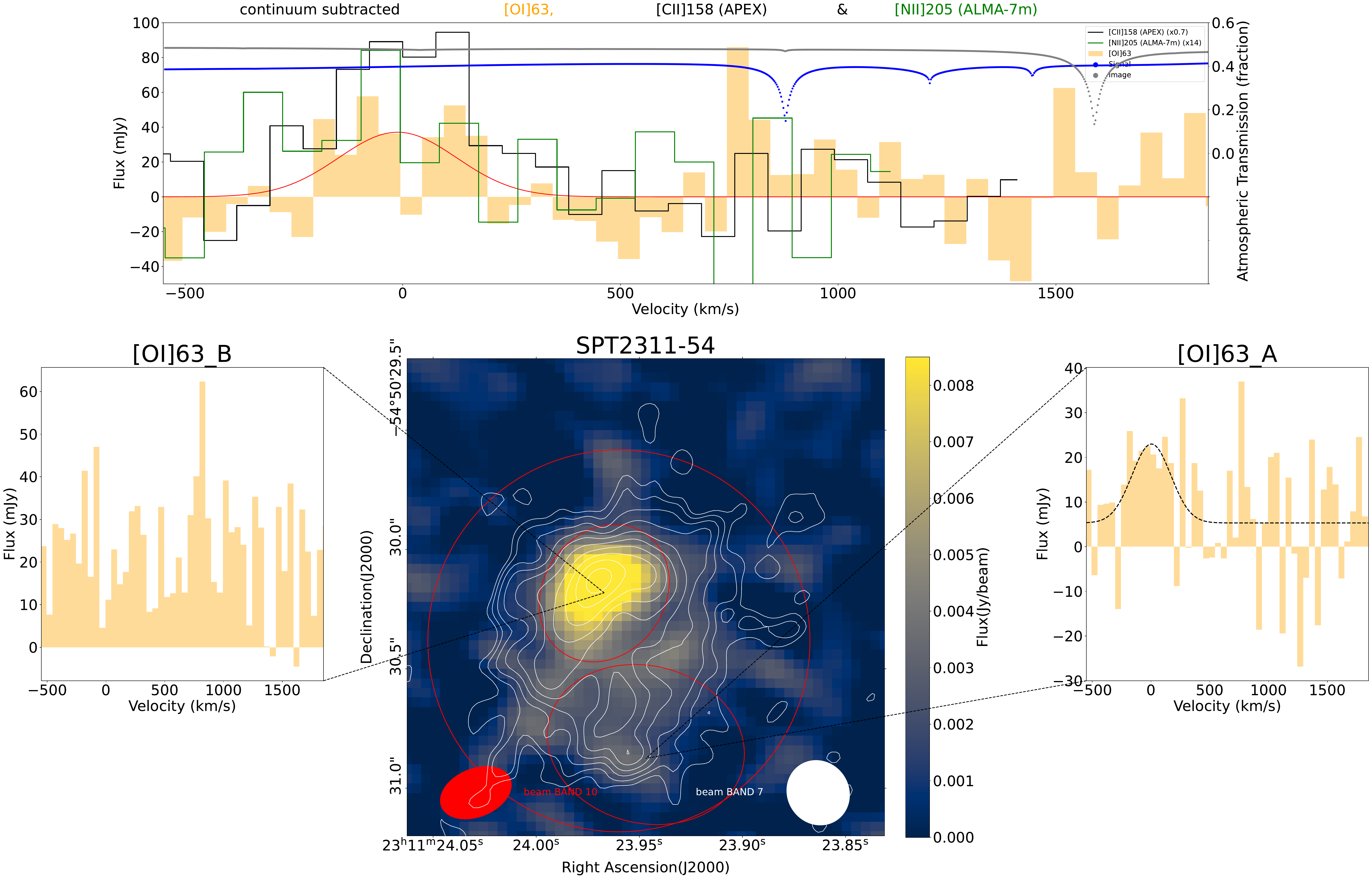}\\
\vspace{0.5cm}
\includegraphics[width=16cm]{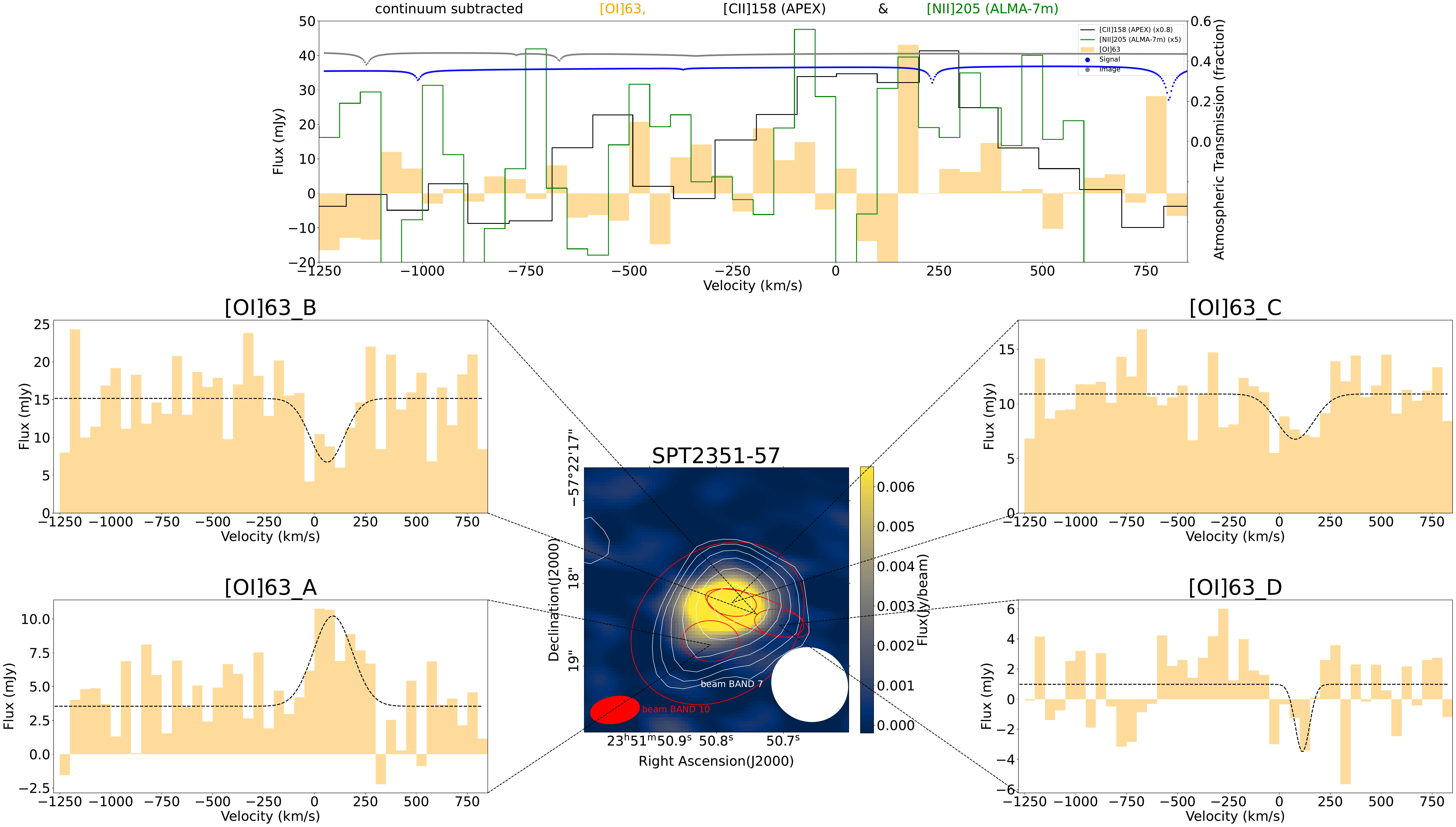}
      \caption{Overview of SPT2311-54 and SPT2351-57 as in  Fig.~\ref{figure:SPT0106OI-overview}.}
      \label{figure:SPT2311OI-overview}
\end{figure*}

The top panels of Figs.\,\ref{figure:SPT0106OI-overview} to \ref{figure:SPT2311OI-overview} show the integrated \OI\ spectra as yellow histograms, with the integrated \CII\ and \NII\ (where available) spectra overplotted to indicate the expected velocity extent of the emission. Four sources also have \OIother\ spectra, which will be reported in a companion paper (Hermans et al., in prep.). The apertures used (listed in column 5 of Table~\ref{table:measurements}) are based on the total extent of the Band 7 and 10 continuum data, which is mostly defined by the Einstein ring morphology. Note that the we cannot use the $\sim$11\arcsec\ APEX beam size as this would extend beyond the Band 10 primary beam.  Our fluxes or upper limits could therefore be considered conservative.

In these integrated spectra, we find only tentative \OI\ detections in two sources (SPT0459-58 and SPT2132-58), with the remaining 10 sources being clear non-detections. 
This generalises the previously reported trend of non-detections \citep{Litke2022,Rybak2023}, but contrasts with the strong detection of \citet{FernandezAranda2024}.
To determine reliable upper limits on our \OI\ fluxes, we estimated the RMS noise by creating moment-0 maps of each source integrated over the \CII\ velocity ranges. Within the Full source aperture (see Table\,\ref{table:measurements}), we randomly placed 1000 beam-sized apertures and measured the flux from the moment-0 map in each. The resulting flux distribution was fitted with a one-dimensional Gaussian, and the standard deviation of this fit was adopted as the RMS noise level.  Table\ref{table:measurements} lists our derived line fluxes and upper limits, together with the \CII\ and \NII\ fluxes with the apertures used.

\begin{figure}
\includegraphics[width=8cm]{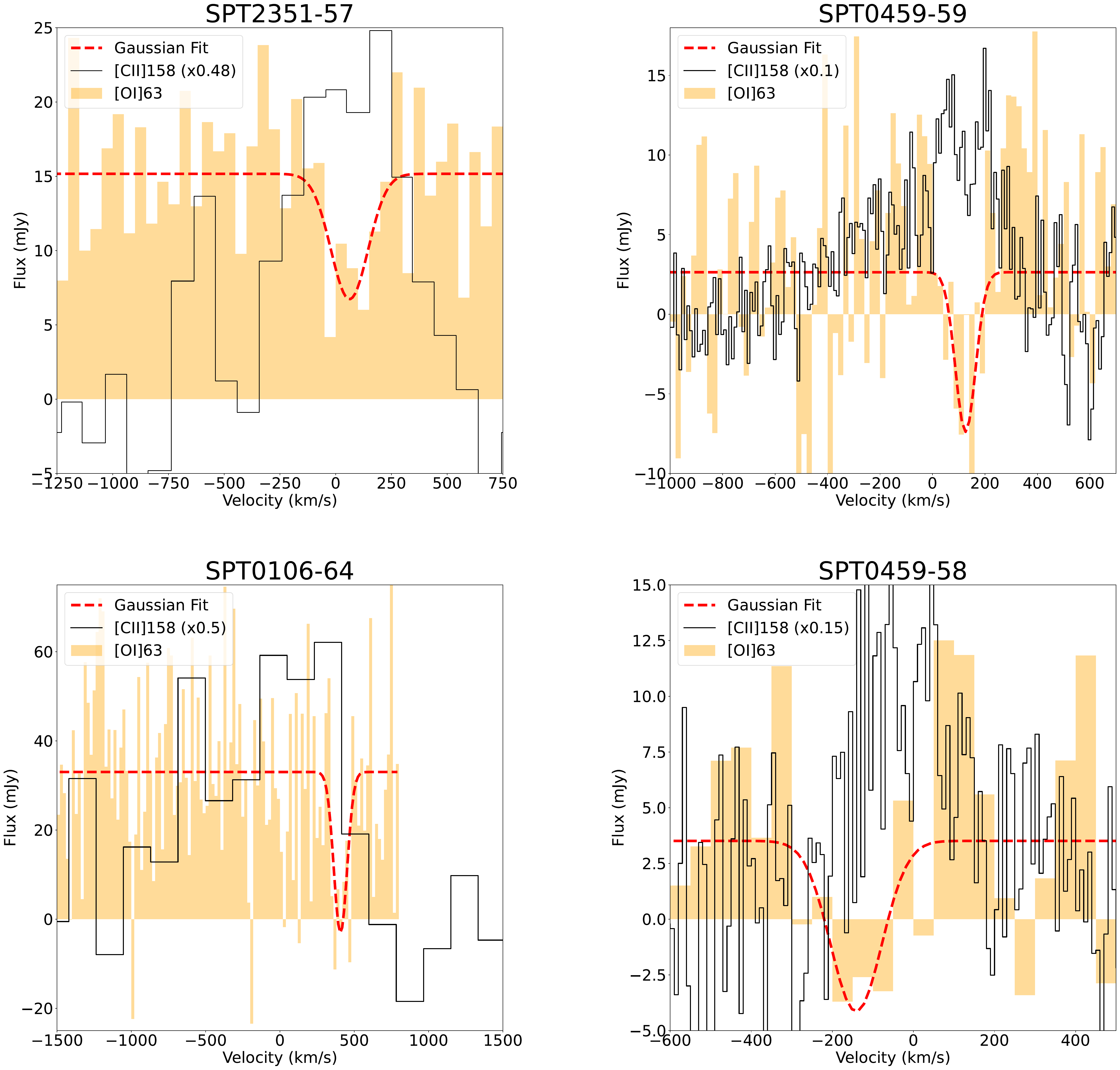}
      \caption{Zoom-in \OI\ spectra of the 4 targets showing tentative absorption against the continuum with integrated \CII\ lines overlaid as black lines. See Figs.~\ref{figure:SPT0106OI-overview} to \ref{figure:SPT2311OI-overview} the full spectra.}
    \label{figure:absorption}
\end{figure}

To benefit from the available high spatial and spectral resolution, we also extracted \OI\ spectra in apertures of particular interest, as described per source in Appendix~\ref{section:individual-sources}. The side panels of Figs.~\ref{figure:SPT0106OI-overview} to \ref{figure:SPT2311OI-overview} show these component extractions labeled alphabetically, with their positions, sizes and fluxes listed in Table~\ref{table:measurements}. The general trends can be summarised as follows:
\begin{itemize}
    \item Continuum components are reliably detected, but no \OI\ emission is detected at these locations.
    \item Narrow velocity \OI\ emission is detected only in positions where there is no continuum emission (e.g. SPT0136-63 B, SPT0155-62 B, SPT0418-47 D, SPT0441-46 D, SPT0459-59 A, SPT2146-55 A, SPT2351-57 A).
    \item Several of the brightest continuum components instead show \OI\ in absorption (SPT0106-64 A, SPT0459-58 A, SPT0459-59 B, SPT2351-57 B, see Fig.~\ref{figure:absorption}).
\end{itemize}

Overall, our observations indicate that the \OI\ emission is strongly suppressed, and can only escape the galaxies through narrow velocity and spatial channels. In the following, we refer to these as "escape channels".

\section{Discussion}

Our spatially resolved \OI\ survey quadruples the number of such reported observations in $z$$>$1 galaxies, compared to the previous two detections \citep{FernandezAranda2024,Ishii2025} and two non-detections \citep{Litke2022,Rybak2023}. It also follows the trend of low \OI\ detection rates reported from {\it Herschel} observations of $z$$>$1 galaxies \citep[e.g.][]{Sturm2010,Coppin2012,Brisbin2015,Wardlow2017,Zhang2018,Wagg2020}, but contrasts with the fact that \OI{} has been observed to be among the top three most luminous FSL in nearby galaxies \citep[e.g.][]{Brauher2008,madden2013,cormier2015,fernandezontiveros2016}. Our new results show that the low spatial and/or spectral resolution of {\it Herschel} and APEX leads to the faint {\it integrated} \OI\ fluxes in those observations, as the galaxy-integrated measurements average over both absorbed and unabsorbed regions. Similarly, stacking galaxy-integrated spectra may lead to a detection when adding up a sufficient number of "escape channels" in these sources. However, our results show that high spatial resolution measurements are required to avoid galaxy-wide measurements which are biased towards the intrinsically small emitting regions.
We now place our \OI\ survey results in context.

\subsection{Line ratio plots}\label{section:line-ratio-plots}

\begin{figure}[ht]
    \includegraphics[width=8cm]{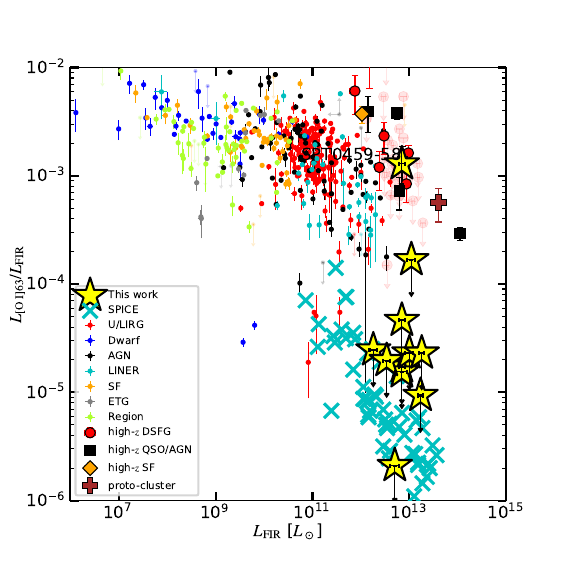}
      \caption{$L_{\rm [OI]63}/L_{\rm FIR}$ vs. $L_{\rm FIR}$ with the 12 SPT DSFGs shown as yellow stars. The literature data are taken from FLAMES \citep{peng2025b}, with references listed in Appendix~\ref{section:data-origin}. The SPT DSFGs are up to $>$100$\times$ fainter than the literature counterparts. The cyan crosses show the SPICE simulated data (\S~\ref{section:simulations}), which are very consistent with the observed SPT DFSGs.}
    \label{figure:OIdeficit}
\end{figure}

Foremost, it is important to verify if our \OI\ observations reached a sufficient depth to draw meaningful conclusions from our mostly non-detections. Fig.~\ref{figure:OIdeficit} adds our 12 datapoints on the compendium of both low-$z$ and high-$z$ galaxies of \citet{peng2025b}. We find that our \OI\ non-detections are 1--2 orders of magnitude deeper than previous detections. The \OI\ emission in our $z$$>$4 DSFGs is therefore either intrinsically less luminous or the emission is strongly absorbed. In the former interpretation, we could expect to find similar trends in some of the other lines, but the \CII\ and even the \NII\ (which traces HII regions rather than PDRs) are fully consistent with the other samples (Fig.~\ref{figure:NIIratio}). Furthermore, the \OIother{} line is observed to be sometimes more than a magnitude more luminous than \OI{} in SPT0418-47 \citep{debreuck2019} and three other sources from our sample (Hermans et al., in prep.), while most models predict this line to be rather $\sim$10$\times$ fainter \citep[e.g.][]{Goldsmith2019,Parente2026}. As the observed extreme \OIshort{} line ratios would require an unphysical interpretation, we therefore conclude that the \OI\ must be strongly absorbed.
Any studies interpreting the ratio of \OI{} with other line or FIR luminosity therefore need to carefully take this absorption into account. 

Two conclusions can be drawn from the $L_{\rm [OI]63}/L_{\rm FIR}$ vs. $L_{\rm FIR}$ plot (Fig.~\ref{figure:OIdeficit}). First, the two previously detected \OI\ lines are in AGN-dominated sources W2246-0526 \citep{FernandezAranda2024} and J2054-0005 \citep{Ishii2025}. While the statistics are still low, this suggests either an enhanced \OI\ emission \citep[e.g.][]{Abel2009,Richings2021} or a reduced absorption in AGN-dominated sources, possibly as a result of higher global temperatures and less amounts of cold, low-excitation \OI{} absorbing material. Second, the $L_{\rm [OI]63}/L_{\rm FIR}$ values below 10$^{-4}$ observed in the SPT DSFGs are rarely seen in lower redshift galaxies. One caveat is that many of the {\it Herschel} observations at 1$<$$z$$<$3 are non-detections, and could have similar ratios if observed to similar depths as our ALMA data. The observed trend may thus still be with FIR luminosity rather than with redshifts. Sensitive ALMA \OI\ observations of less luminous FIR sources at $z$$>$4 are required to elucidate this.

\begin{figure}[ht]
    \includegraphics[width=8cm]{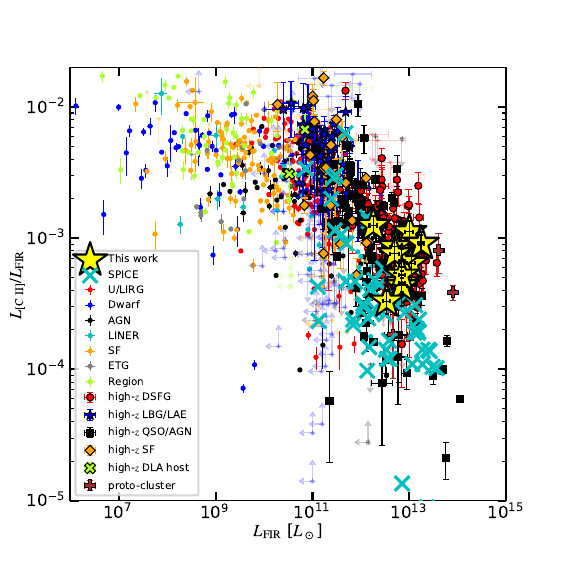}
      \caption{$L_{\rm [CII]}/L_{\rm FIR}$ vs. $L_{\rm FIR}$ with the 12 SPT DSFGs shown as yellow stars. The literature data are taken from FLAMES \citep{peng2025b}, with references listed in Appendix~\ref{section:data-origin}. The SPT DSFGs follow the general population. The cyan crosses show the SPICE simulated data (\S~\ref{section:simulations}), which are slightly under-predicting the $L_{\rm [CII]}/L_{\rm FIR}$.}
      \label{figure:CIIdeficit}
\end{figure}

\begin{figure}[ht]
    \includegraphics[width=8cm]{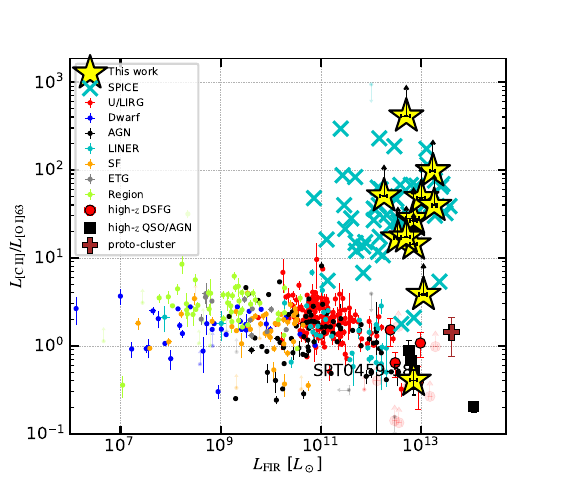}
      \caption{$L_{\rm [CII]}/L_{\rm [OI]63}$ vs. $L_{\rm FIR}$ with the 12 SPT DSFGs shown as yellow stars. The literature data are taken from FLAMES \citep{peng2025b}, with references listed in Appendix~\ref{section:data-origin}. The SPT DSFGs are up to $>$100$\times$ fainter in \OI{} than the literature counterparts. The cyan crosses show the SPICE simulated data (\S~\ref{section:simulations}), which are slightly lower than the observed SPT DFSGs.}
      \label{figure:CII-OI63 ratio}
\end{figure}

\begin{figure}[ht]
    \includegraphics[width=8cm]{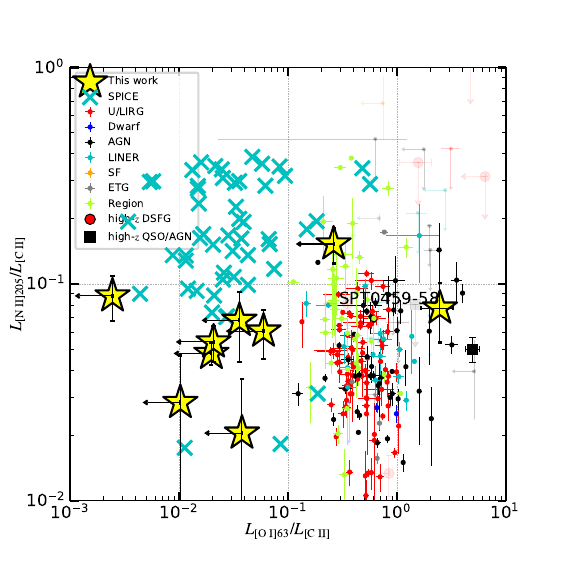}
      \caption{$L_{\rm [NII]205}/L_{\rm [CII]}$ vs. $L_{\rm [OI]63}/L_{\rm [CII]}$ with the 9 SPT DSFGs shown as yellow stars. The literature data are taken from FLAMES \citep{peng2025b}, with references listed in Appendix~\ref{section:data-origin}. As in Fig.~\ref{figure:CIIdeficit}, the SPT DSFGs are consistent with the literature counterparts in $L_{\rm [NII]205}/L_{\rm [CII]}$, while they are offset in $L_{\rm [OI]63}/L_{\rm [CII]}$. The cyan crosses show the SPICE simulated data (\S~\ref{section:simulations}), which are slightly higher in $L_{\rm [NII]205}/L_{\rm [CII]}$ than the observed SPT DFSGs, which may be a combination of over-predicted \CII{} and over-predicted \NII.}
      \label{figure:NIIratio}
\end{figure}

\subsection{The nature of the strong \texorpdfstring{\OI{}}{[OI] 63 micron} absorbers}\label{section:absorbers}

\subsubsection{Comparison with Galactic and lower redshift results}
Absorption features in the \OI\ line and against the dust continuum were already reported in the first spectrally resolved \OI\ observations targeting the DR21 star-forming region \citep{Poglitsch1996} and the main core of the Sgr B2 complex \citep{Baluteau1997,Vastel2002}. This can be explained by a high column density of atomic oxygen of this optically thick line, creating absorption by O$^0$ either within the emitting cloud itself, or by foreground clouds within the ISM along the line of sight. 
Already for column densities $N_{\rm H}\gtrsim 10^{22}$\,cm$^{-2}$, which are quite common in the neutral medium, the \OI\ line becomes optically thick for line widths of 10\,\kms \citep{Decarli2025}. In contrast, the \OIother\ line will almost always be optically thin, but the \CII\ line could also become optically thick at $N_{\rm H}\gtrsim$ 4$\times 10^{22}$\,cm$^{-2}$, and has shown similar, tough shallower, absorption profiles \citep[e.g.][]{Guevara2020}.

Spatially and spectrally resolved data can distinguish between the origin of the absorption within the emitting or foreground cloud, as the latter will often have an offset in velocity. Such observations have been difficult as 63\,$\mu$m observations can only be done with relatively small space telescopes ({\it Infrared Space Observatory, Herschel}) or airborne telescopes (Kuiper Airborne Observatory and Stratospheric Observatory For Infrared Astronomy (SOFIA)). Only in several nearby Galactic star-forming regions have such resolved observations been possible \citep[e.g.][]{Kraemer1998,Leurini2015,Gerin2015,Gusdorf2017,Schneider2018,Mookerjea2019,Goldsmith2021,Jackson2024}. These studies favoured the interpretation that the \OI\ absorption originates predominantly in foreground clouds within the ISM, as many regions show  clear absorption profiles not only against the emission profile, but also against the background continuum. From a comparison of the \OI, $^{13}$CO and HI absorption profiles, \citet{Lis2001} demonstrated that, although regions are expected to be primarily molecular, the O$^0$ abundance is non-negligible, with a linear relationship suggesting a correlation with CO gas column density and a gas-phase abundance of neutral Oxygen that is about 10\% of CO.
The regions with strong [OI] line emission are found to have a factor $\sim$3 reduction in the spectro-spatially integrated [OI]63 intensity, as geometric effects (such as HII regions and incident UV flux density existing behind lines-of-sight to cold foreground clouds) influences the observed emission \citep{Goldsmith2021}.
Thanks to the improved spectral resolution of SOFIA, these absorbers could be studied in multiple tracers such as the optically thin $^{13}$\CIIshort\ and \OIother, showing that the absorbing O$^0$ is co-spatial with C$^+$ in a dense, low excitation foreground cloud associated with the PDR \citep{Guevara2024}. One prediction from these Galactic studies is thus that we may see a shallower version of these absorption profiles in other FSL such as \CII, provided they are  observed at at least comparable spatial and spectral resolution.

In low-redshift galaxies, only low spectral resolution and spatially unresolved 63\,$\mu$m observations have been possible \citep{rosenberg_2015,Israel2017,Kramer2020}. These studies all mention the possible impact of \OI\ self-absorption, but only the intrinsic absorption within the clouds themselves can be considered within PDR models, as the foreground absorbing component with likely very different excitation conditions cannot be easily separated within the limited resolution element of the 3.5\,m {\it Herschel} telescope. The observed ratio of the 63 to 145$\mu$m \OIshort\ lines is close to the predicted values from PDR models, implying that the amount of \OI\ absorption should be relatively modest in these nearby galaxy samples \citep[e.g.][]{fernandezontiveros2016}. For those sources containing an AGN, the \OI\ flux may also have been boosted by the presence of X-ray dominated regions \citep[e.g.][]{Maloney1996,Dale2004,Brauher2008,Herrera2018}. However, observational evidence of velocity-resolved \OI\ absorption has been seen in several {\it Herschel} spectra of (ultra-)luminous infrared galaxies \citep[(U)LIRGs;][]{GonzalezAlfonso2012,rosenberg_2015,diazsantos2017}. 

Interestingly, the 63$\mu$m observational capabilities are significantly better in $z>4$ galaxies than in all but the nearest Galactic star-forming regions, thanks to the very high spectral and spatial resolution capabilities of ALMA. The main limitation is the velocity spread of the different emitting clouds within the synthesized beam caused by galaxy rotation. Our detection of narrow "escape channels" where we do detect \OI\ like in region D of SPT0418-47 demonstrates that we are now reaching the required spatial and spectral resolution, possibly helped by the strong gravitational magnification in this source\footnote{Amplification $\propto \sqrt\mu \approx 5.7$, with $\mu$=32.7 \citep{Spilker2016}}, allowing to reach spatial scales below 200\,pc. Similarly, the two sources observed at $<$0\farcs05 resolution (SPT2132-58 and SPT2146-55), although having more moderate lensing factors of $\mu\approx 6$, also reach spatial scales 100--150\,pc, and show tentative \OI\ detections in some of the components (Fig.~\ref{figure:SPT2132OI-overview}). Detailed lens modelling of the \OI\ dataset will be done in a future publication to determine the exact scales.

Despite these narrow line detections, the overall conclusion from our \OI\ survey in the SPT DSFGs is that they are affected by 1-2 orders of magnitude stronger absorption than the low-redshift ULIRGs (Fig.~\ref{figure:OIdeficit}). A comparison with \OI{} in high-$z$ DSFGs is difficult as most observations resulted in upper limits or very marginal detections \citep{Coppin2012,Brisbin2015,Wardlow2017,Zhang2018,Wagg2020,Litke2022,Rybak2023,Ishii2025}, with only one solid detection reported in an AGN-dominated source \citep{FernandezAranda2024}. Only three literature high-$z$ DSFGs have both \OI{} and \CII{} observed (Fig.~\ref{figure:CII-OI63 ratio}), showing ratios comparable to the low-$z$ DSFGs, suggesting a real difference with our SPT DSFG observations. However, publication bias against non-detections may strongly bias this small literature sample.  In our statistically significant SPT DSFG \OI{} sample, the low \OI{} luminosity (Fig.~\ref{figure:OIdeficit}) contrasts with their 'normal' \CIIshort{} (Fig.~\ref{figure:CIIdeficit}). 
We have already considered and rejected the possibility that this could be due to fainter intrinsic \OI{} (\S~\ref{section:line-ratio-plots}, Hermans et al., in prep.). To completely suppress the \OI\ emission, these absorbers would need to cover a spread of velocities at least as wide as the emitting gas. Figure~\ref{figure:absorption} zooms in on four strong absorbers; although we are severely limited by both S/N and the fact that only integrated \CII\ profiles are available, in at least two cases, the strongest \OI{} absorber seems to coincide with the peak of the \CII\ emission, suggesting that the absorbers are mostly following the overall kinematics of their host galaxies. The presence of absorption against the continuum e.g. in region A of SPT0106-64 indicates that these are foreground clouds rather than peripheral regions of the emitting clouds.  
A foreground neutral gas density of $N_{\rm H}\gtrsim 10^{22}$\,cm$^{-2}$ is sufficient to cause discernable \OI\ absorption, because at such high column densities $N(O)$ scales with $N(H+H_2)$; with lots of $N(O)$ that is not excited in cold clouds, there is sub-thermally excited \OI{} ready to absorb any emission along the line of sight. Any photons that escaped the higher density, higher temperature and higher optical depth regions where the emission occurs would likely be absorbed along the line of sight, explaining why such absorbing clouds could be quite widespread in the SPT DSFGs. 
At the redshifts of our SPT DSFGs, an observed increase of the dense gas fraction from molecular tracers such as HCN, HNC and HCO$^+$ \citep{Rybak2026} has also been observed, which may explain the difference with local ULIRGs located in regions with a lower ISM density. While most of the oxygen is expected to be in the molecular rather than the atomic form, the overall density of foreground atomic oxygen susceptible to create absorption will likely also increase in these regions with a denser ISM. The strong \OI\ absorption may thus originate from the dense component of the interstellar, intergalactic or even circumgalactic medium of these high-$z$ DSFGs, while in local ULIRGs the star formation is more concentrated in episodical bursts of star-formation in compact central regions \citep[e.g. in Arp220,][]{Martin2016}, with less  remaining dense gas along the line of sight. 

\subsubsection{Implications}
The implications of the strong \OI\ absorption are multiple. First, deep absorption features appear in some systems to have been measured in contrast to the level of the Cosmic Microwave Background (CMB) radiation (i.e. down to flux densities equal to 0 or below, see Fig.~\ref{figure:absorption}). The absorption troughs of this ground-state line transition are centred on the redshifted line frequency and appear as coincident with \CIIshort{} line features at the same velocities, see e.g. SPT0459-59. The CMB radiation temperature is $\sim$16K at a fiducial $z$ = 4.8. This low background temperature may yield absorption features in contrast to the CMB radiation if enough sub-thermally excited Oxygen atoms in the ground state have a low enough excitation temperature to absorb the CMB photons. In the optically thin limit, with a fiducial gas kinetic temperature = 100\,K, excitation temperatures reach $T_{\rm exc}$$\sim$25\,K for both collisions with H and H$_2$ for \OI{} at densities of $\leq$ 100\,\percc\, and further down to 14.9\,K at $\leq$ 1\percc{} \citep{Goldsmith2019}. \OI{} is sub-thermally excited in low-density conditions ($n(H)$$<$$n_{\rm critical}$). 
Such conditions may exist in well-shielded, cold dust (10-15 K) in low-density clouds, for which these lines-of-sight can result in such absorption against the CMB radiation \citep{Riechers2022}. Due to the aperture-dependence of this absorption feature and the limited signal-to-noise, the evidence for absorption of \OI{} in contrast to the redshifted CMB radiation will require a more systematic investigation with deeper integration times, ideally with additional information from \OIother{} transition, to conduct a more robust radiative transfer analyses to disentangle these potential \OI{} absorption line features against the CMB radiation.

Second, the strength of the absorption makes it virtually impossible to recover the intrinsic \OI\ flux integrated over the entire. Although the \OIother\ line is predicted to be $\sim$10$\times$ fainter, it is a better choice for statistical studies of intrinsic line ratios involving the \OIshort\ lines \citep[e.g.][]{Lee2021,debreuck2019,Fudamoto2025}. 

Third, for a given column density, the \CIIshort\ line has an optical depth 4$\times$ lower than that of \OI\ \citep{Decarli2025}, meaning that for $\tau$(\OI{})$>$4, the \CII\ would also become optically thick. Given that in some sources $>$99\% of the expected \OI\ flux is absorbed, $\tau$(\OI)$\gtrsim$4.5, and $\tau$(\CIIshort)$\gtrsim$1.1. While this estimate is very rough and depends on other effects such as the C to O abundance ratio, it illustrates that optical depth effects in \CIIshort\ cannot be neglected \citep[see also][]{gullberg2015}. The contributions from the ionised gas phase to the \CIIshort\ line would dilute the \CIIshort\ self-absorption signal, but we expect these to be relatively modest given the relatively low \NIIlong/\CIIshort\ ratios in the 12 sources in our sample (Fig.~\ref{figure:NIIratio}) suggesting that the neutral gas dominates the \CIIshort\ emission \citep{Decarli2014,Cunningham2020}. 
Additionally, as \CII\ traces less dense and more diffuse gas spread over much larger scales throughout the galaxies, the foreground \CII-absorbing gas would need to be even more widespread compared to \OIshort; any \CII\ absorption would therefore likely be concentrated in smaller regions where the optical depth of the foreground material is higher. 
A tentative detection of \CII\ absorption can be seen in SPT0459-59 (see Fig.~\ref{figure:absorption}), which intriguingly coincides in velocity with the strongest dip in the \OI{}, quite reminiscent to what is seen in Mon R2 and M17 SW \citep{Guevara2024}. Deeper \OI\ data and matched spatial resolution \CII\ observations are required to determine the impact of optical depth effects on the \CII{} emission. 
However, the fact that our strongly absorbed \OI\ sources follow the overall population in \CIIshort{} vs. FIR (Fig.~\ref{figure:CIIdeficit}) suggests this is a modest effect when integrated over the full galaxy.

Fourth, absorbing gas is a powerful tracer of outflow (or inflow) kinematics, which has previously only been used in molecular lines such as OH \citep[e.g.][]{Spilker2020}; such studies can now also be considered for the more abundant atomic oxygen by looking for (inverse) P-Cygni profiles \citep[e.g.][]{Ishii2025}.

Fifth, the fact that the foreground absorbing material is rich in oxygen implies it consists of chemically enriched rather than primordial material. This may argue for the absorbers to be located within the galaxies, though absorbed ionised metal lines are regularly found in restframe UV spectroscopy of $z$$\sim$3 radio galaxies \citep[e.g.][]{Kolwa2019}.

\subsection{The role of dust continuum?}

In addition to the above result that foreground clouds in the ISM can (almost) completely wipe out the \OI{} emission in the SPT DSFGs, we also find that the narrow \OI{} "escape channels" seem to occur only in regions without any dust continuum emission. This is most evident in SPT0418-47 (see Fig.~\ref{figure:SPT0418OI-overview}), where our strongest \OI{} detection is located in the dust-free region D. To understand the origin of this effect, it is important to have an idea of the optical depth of the dust emission at $\lambda_{\rm rest}$=63\,$\mu$m. 
To determine the optical depth requires photometric coverage on either side of the dust continuum peak, which is generally only possible with spatially unresolved {\it Herschel} data covering the Wien side. The SPT DFSG sample has such coverage, and the peak is found to between $\lambda_{\rm rest}$=66 and 154\,$\mu$m \citep{Reuter2020}. 
It it therefore likely that the dust opacity in the SPT DSFGs is non-negligible at $\lambda_{\rm rest}$=63\,$\mu$m, and may significantly attenuate the \OI{} line brightness when averaged across each system. One example is the \textit{Planck-Herschel} selected Red Radio Ring at $z$=2.55, with $\lambda_{\rm 0}\simeq$200\,$\mu$m \citep{Harrington2019}. A foreground dust screen (i.e. a factor $e^{\tau_{205}}$) with a molecular gas column density $>$10$^{24}$\,cm$^{-2}$, and molecular gas to dust mass ratio (GDMR) of 100, is found to reduce even the \NIIlong\ line by a factor of $\sim$5 \citep{Harrington2019}; extrapolating this to \OI{} would imply a reduction by 100$\times$ or higher for such a simplified dust screen. 
However, we note that values derived in \citet{Reuter2020},  \citet{Harrington2019}, and other studies, are often calculating the value at which dust becomes optically thick based a single dust component. The dust that is optically thick at 100\micron\ may be biased towards higher column density regions, which would in turn bias the derived dust opacities averaged over an entire source -- even when the small source solid angle emitting regions with the highest densities and dust optical depths are a small fraction of the source. The solution is to obtain spatially resolved studies of the dust SEDs, which is becoming possible at $z$$\gtrsim$4 when the peak shifts through ALMA bands 10 and 9 \citep{FernandezAranda2025,Meyer2025,Alegera2025}. While there is a significant source-to-source variation, these studies find an increase of the optical depth towards the central regions of the galaxies, where the dust becomes optically thick around $\lambda_{\rm rest}$=100\,$\mu$m. Our Band 10 continuum data of these 12 SPT DSFGs thus provides a powerful spatially resolved study of this dust continuum into the optically thick regime down to spatial resolutions of $\sim$100\,pc in the source plane. Several SPT DSFGs have sub-arcsecond observations covering a range of wavelengths (see e.g. the Band 7 data in Figs.~\ref{figure:SPT0106OI-overview} to \ref{figure:SPT2311OI-overview}), which will allow to determine the spatial variation of the dust opacity and compare them with our \OI{} data. We defer such a detailed study to future publications.

While it is clear that attenuation by optically thick dust may reduce the observed \OI{} flux, this cannot fully explain our results. First, dust attenuation would impact the full line profile, but we find cases where only part of the lines are absorbed, e.g. the narrow velocity absorbers against the continuum (see Fig.~\ref{figure:absorption}). 
Second, optically thick dust does not explain the preferential location of the \OI{} escape channels in areas of low dust continuum. 
A more straightforward interpretation is that at least part of the spatially resolved dust emission is optically thin. The observed continuum emission would then be a direct tracer of the dust mass along the line of sight. Assuming a constant gas/dust ratio, this would imply less absorbing gas in the locations of lower dust emission, creating the \OI{} escape channels. The low temperatures of cold dust-rich molecular clouds of 20-50 K would also minimally impact the observed brightness temperatures of \OI{} \citep{Goldsmith2021}, leaving sub-thermally excited, oxygen-rich foreground material dispersed within the ISM to absorb the ground-state \OI{} line emission. Together, we conclude that the role of dust optical depth from foreground dust is less important than high line opacities in \OI{} emitting regions and the foreground cold molecular clouds dispersed throughout the DSFGs, the latter with large amounts of absorbing material that is sub-thermally excited and has high enough oxygen column density.   

\subsection{Comparison with simulations}\label{section:simulations}
Modelling oxygen lines in cosmological simulations remains challenging due to strong self-absorption and dependence on ISM modelling. Previous studies predict \OI~observables \citep{RamosPadilla2023}, albeit without the inclusion of self-absorption effects. More recently, \citet{Parente2026} found foreground self-absorption can reduce the \OI{} flux by a factor of 2-4. However, for ISM sensitive lines such as \OI{}, resolving the multiphase ISM is key. 
In this work we employ {\tt SPICE}, a suite of cosmological radiation-hydrodynamical (RHD) simulations \citep{Bhagwat2024}. These simulations are performed with {\sc ramses-rt}   \citep{Rosdahl_2013,Rosdahl_2015}, the coupled RHD extension of the Eulerian, adaptive mesh
refinement code {\sc ramses} \citep{Teyssier_2002}. The key advantage of the {\tt SPICE} simulations is their self-consistent tracking of radiation in 5 frequency bins including the Habing and ionising bands which are essential to model \CIIshort{} and \OI{} lines in addition to $\sim30-60 \ \rm pc$ resolution in the ISM/CGM. {\tt SPICE} follows the non-equilibrium chemistry which is fully coupled to the local radiation field for $e^-$, H\,{\sc{i}}, He\,{\sc{ii}} and He\,{\sc{iii}}. Star formation in {\tt SPICE} incorporates variable efficiencies within the ISM that respond to the local turbulent state of the gas \citep{Kretschmer_2020}. Feedback from these stars is injected in the form of radiative feedback (photo-ionisation and -heating) and radiation pressure on dust. Additionally, supernovae inject mass, metals (tracked and advected as passive scalars) and momentum back into the ISM driving large scale galactic outflows (see \citet{Bhagwat2024} for details). Dust in {\tt SPICE} is coupled to H\,{\sc{i}} chemistry such that the dust density is defined as $n_d = \frac{Z}{Z_\odot} f_{\rm HI} n_{\rm gas}$, where $Z$ is the metallicity, $f_{\rm HI}$ is the H\,{\sc{i}} fraction and $n_{\rm gas}$ is the gas density. We use the $\approx15$ Mpc volume of {\tt SPICE} with a dark matter and baryonic resolution of $\approx6.3\times10^5 \ M_\odot$ and $\approx975 \ M_\odot$. 
Finally, we select the more massive SPICE galaxies with
$10^{9} \lesssim M_\star \lesssim 2\times10^{10}\,M_\odot$
at $z = 5$--$6$, yielding a sample of 48 objects. This selection probes
the low-mass regime of DSFGs and overlaps with the intrinsic stellar masses inferred for the SPT sources \citep{Cathey2024}. Moreover, the mock galaxies reach near-solar metallicities, consistent with the metallicity estimate reported for SPT0418-47 \citep{debreuck2019}.

We model \OI{}, \CIIshort{} and \NII{} lines by assigning an emissivity to each gas cell in the simulation outputs (for details see Bhagwat et al. in prep.) as
\begin{equation}
    L_{\rm ion} =  \sum^{e^-,\rm HI}_i n_i n_{\rm ion}, \Lambda_i (T) V,
    \label{eq:Lum}
\end{equation}
For all lines, we also calculate the self-absorption by assigning an optical depth per cell and estimating an angle averaged attenuation for 100 lines of sight out to the virial radius of each halo. Optical depth is defined as 
\begin{equation}
    \tau_{\nu_0} = \frac{c^3}{8 \pi \nu_0^3}A_{\rm ul} \frac{g_l}{Z(T_{\rm ex})} N_{\rm ion} \biggl[1 - {\rm exp}\biggl(-\frac{h\nu_0}{k T_{\rm ex}} \biggr)\biggr] \frac{1}{\Delta \nu},
\end{equation}
where $\nu_0$ is the rest frequency of given line, $A_{\rm ul}$ is the spontaneous emission coefficient, $g_l$ and $g_u$ are the level degeneracies, $\mathcal{Z}(T_{\rm ex}) = g_l + g_u {\rm exp}(-91.2/T_{\rm ex})$ is the partition function (in the case of \CII), $N_{ion}$ is the column density of ion in consideration, $\Delta \nu$ is the Doppler width and $T_{\rm ex}$ is the excitation temperature for the transition. All line ratios and luminosities presented in this work are integrated over the galaxy ignoring instrumental effects.

The results from the simulations are plotted as cyan crosses in Figs.~\ref{figure:CIIdeficit}, \ref{figure:OIdeficit}, \ref{figure:CII-OI63 ratio}, and \ref{figure:NIIratio}. The SPICE data slightly under-predict the \CIIshort{} luminosities and have a large scatter on the \NII, which may originate from the very different ionisation states of gas from the different feedback in
SPICE. However, the model predictions are close to the highly absorbed \OI{} luminosities (Fig.~\ref{figure:OIdeficit}), suggesting that we should detect the emission with integrations reaching 2--10$\times$ lower rms than our relatively short integrations of 15-30\,minutes. Overall, this comparison highlights the critical importance of self-absorption when making predictions of optically thick lines such as \OI{}.


\section{Conclusions}
We have carried out the first dedicated ALMA survey of the \OI{} fine-structure line in a relatively large sample of 12 gravitationally lensed DSFGs at $z>4$. Our main conclusions are as follows:

\begin{enumerate}
    \item Despite reaching sensitivities significantly deeper than previous high-redshift studies, we detect only tentative \OI{} emission in two out of twelve sources. The remaining galaxies show non-detections that extend the dynamic range of existing constraints by one to two orders of magnitude after strong gravitational lensing considerations.
    
    \item The low \OI{} luminosities cannot be readily attributed to intrinsically weak line emission, as other far-infrared fine-structure lines, such as \CII{} and \NII{}, and in four cases also \OIother\ are detected at levels consistent with comparable galaxy populations. The observations instead imply substantial absorption of the \OI{} line by low-excitation neutral gas within widespread molecular clouds in these DSFGs, together with high line optical depths in the regions where \OI{} emission occurs.
        
    \item The anti-correlation between \OI{} emission and the brightest dust continuum regions suggests that line absorption effects, rather than possibly high dust optical depths at 63\micron, are primarily responsible for suppressing the observed line emission. This can explain the low \OI{} detection rate in the SPT DSFG, which has been selected by its cold dust emission at mm wavelengths \citep{vieira2010}. As a corollary, differently selected galaxies may have a higher \OI{} detection rate.
    Future resolved dust SED modelling will further inform the role of infrared radiation field and dust continuum effects for these 100-150 pc-sized regions. 
    
    \item Several sources exhibit \OI{} absorption against the dust continuum, while others show narrow ($\sim$20\kms), spatially offset emission components. These signatures add observational evidence to indicate that the \OI{} line is subject to strong optical depth and kinematically distinct self-absorption effects. Although S/N is limited, several absorption troughs appear to reach levels below the CMB. This suggests the presence of low-excitation temperature (Tex $\leq$ $T_{\rm CMB}$), low-density gas along those lines of sight.
    
    \item These high-resolution observations suggest that galaxy-integrated \OI{} measurements in high-redshift DSFGs cannot reliably trace the intrinsic cooling rate or density of the neutral ISM. Together with the optically thinner tracer, such as the \OIshort{}\,145\,$\mu$m line, it may be possible to disentangle some of the more complicated radiative transfer effects.
\end{enumerate}

In summary, our results show that the \OI{} line is strongly affected by absorption in $z>4$ DSFGs, limiting its diagnostic power when interpreted in isolation. Future observations combining matched-resolution spectroscopy, spatially resolved dust continuum measurements, and detailed lens modelling will be essential to disentangle intrinsic emission from absorption effects, and to fully exploit the diagnostic potential of far-infrared fine-structure lines at high redshift to recover the underlying physical conditions of the interstellar medium of active galaxies in the early Universe.

\begin{acknowledgements}
We thank Juan Pablo P\'erez Beaupuits for stimulating discussions linking our high redshift results with very similar Galactic data, and Jarle Brinchman for useful suggestions. We thank for the anonymous referee for a very careful reading which allowed us to better emphasise several unexpected results from our observations.
This paper makes use of the following ALMA data: ADS/JAO.ALMA\#2024.0.01465.S, 2021.1.00857.S, 2016.1.00231.S,  2019.1.01026.S, 2021.1.00252.S, 2019.1.00253.S, 2015.1.00942.S, and  2018.1.00046.S. ALMA is a partnership of ESO (representing its member states), NSF (USA) and NINS (Japan), together with NRC (Canada), NSTC and ASIAA (Taiwan), and KASI (Republic of Korea), in cooperation with the Republic of Chile. The Joint ALMA Observatory is operated by ESO, AUI/NRAO and NAOJ. 
IDL acknowledges support through funding from the European Research Council (ERC) under the European Union’s Horizon 2020 research and innovation program DustOrigin (ERC-2019-StG-851622), and from the Flemish Fund for Scientific Research (FWO-Vlaanderen) through the research project G0A1523N.
\end{acknowledgements}

\bibliographystyle{bibtex/aa}
\bibliography{bibtex/OI,bibtex/flames-low_v0.1,bibtex/flames-high_v0.1}

\begin{appendix}
\section{Literature data origin}\label{section:data-origin}
The plotted low-$z$ data are taken from  FLAMES-low catalogue in \citet{peng2025b}, sourced from, SHINING \citep{Herrera2018}, DGS \citep{M13,cormier2015,C19}, HERCULES \citep{rosenberg_2015}, HERUS \citep{Farrah2013}, \citet{diazsantos2017}, \citet{fernandezontiveros2016}, \citet{Z16}, \citet{Brauher2008}, \citet{F14}, \citet{H10}, \citet{K16}, \citet{P21}, \citet{S22}, \citet{Spinoglio2015}, \citet{D15}, \citet{L17a}, \citet{L17b}, \citet{S19}. The plotted high-$z$ data are taken from FLAMES-high catalogue in \citet{peng2025b}, sourced from \citet{TK22}, \citet{Reuter2020}, \citet{FernandezAranda2025}, \citet{Wagg2020}, \citet{WJ13}, \citet{Aravena2016}, \citet{Rybak2020}, \citet{Zhang2018}, \citet{KS25}, \citet{Litke2022}, \citet{DT16}, \citet{SA13}, \citet{ZJ18}, \citet{MG11}, \citet{BC06}, \citet{UB16}, \citet{FernandezAranda2024}, \citet{BA06}, \citet{LD07}, \citet{Brauher2008}, \citet{PP15}, \citet{DT18}, \citet{AM08}, \citet{GM23}, \citet{LC19}, \citet{Sturm2010}, \citet{MG14}, \citet{Ishii2025}, \citet{IR10a}, \citet{FC15}, \citet{Cunningham2020}, \citet{WA07}, \citet{Wardlow2017}, \citet{IR13}, \citet{RM23}, \citet{DT21}, \citet{LK19}, \citet{HS10}, \citet{VI11}, \citet{MH14}, \citet{stacey2010}, \citet{WA03}, \citet{VB20}, \citet{SH18}, \citet{WR13}, \citet{Coppin2012}, \citet{IR16}, \citet{RD18}, \citet{LC18}, \citet{MG12}, \citet{RM20a}, \citet{CA11}, \citet{DR18}, \citet{gullberg2015}, \citet{DD99b}, \citet{RC25}.

\section{Flux calibration}\label{section:flux-calibration}

As described in \S\,\ref{section:Band10observations}, the photometric accuracy of high frequency observations is affected by several uncertainties which generally lead to a lower observed flux than reality. We here compare the observed line+continuum fluxes (which given the strong self-absorption are expected to be dominated by continuum) with published fluxes from {\it Herschel}. We collected both the published fluxes obtained with {\tt HIPE v9.0} \citep{Reuter2020} as well as those derived from the SPIRE point source catalogue\footnote{Available from \url{https://irsa.ipac.caltech.edu/cgi-bin/Gator/nph-scan?submit=Select&projshort=HERSCHEL} and }. The latter fluxes are derived using the timeline fitter method\footnote{See \url{https://irsa.ipac.caltech.edu/data/Herschel/SPSC/docs/SPSC_ExplSup_v1.pdf}}.

Table~\ref{table:alma-spire-comp} presents these fluxes. We find that there are very good matches in five cases (SPT0243-49, SPT0441-46, SPT0459-58, SPT0459-59, and SPT2132-58), while in six sources only 40--80\% of the {\it Herschel} flux is recovered. This discrepancy could be partially due to the flux decoherence effects described in \S\,\ref{section:high-frequency}, but we do not find a clear trend with higher phase rms or larger distance to the calibrators. As the {\it Herschel}/SPIRE beam is 25\arcsec{} or 35\arcsec{} at 350\micron{} or 500\micron, respectively, it is also possible that nearby companions were included, which are not covered in the ALMA apertures of 0\farcs8 to 6\arcsec{} size. We therefore conclude that while remaining absolute flux calibration uncertainties remain in these high frequency data, this does not appear to be a systematic trend, as half of our fluxes are consistent with those from {\it Herschel}.

\begin{table}[h]
\small
\caption{Source Photometry: ALMA compared to Herschel/SPIRE.}
\begin{tabular}{ccccccc}
\hline \hline
Source  & $\lambda_{\rm obs}$ & Flux$^a$  & SPIRE 350  & SPIRE 500 & SPIRE 350 & SPIRE 500 \\
Name & \micron{} & mJy & PSC$^b$ [mJy] & PSC$^b$ [mJy] & Reuter2020 & Reuter2020 \\
\hline
\textbf{SPT0106-64} & 369 & 229$\pm$46 & 271.9 $\pm$ 12 & 276.6$\pm$14 & 256$\pm$10 & 237$\pm$9 \\
\textbf{SPT0136-63} & 338 &  51$\pm$10 & 95.4 $\pm$ 9 & 128.4$\pm$7 & 81$\pm$7 & 122$\pm$6 \\
\textbf{SPT0155-62} & 341 &  50$\pm$10 & ... & ... & 135$\pm$7 & 200$\pm$7 \\
\textbf{SPT0243-49} & 430  & 55$\pm$11 & 40.5 $\pm$ 10 & 63.4 $\pm$ 11 & 44$\pm$7 & 58$\pm$7 \\
\textbf{SPT0418-47} & 333 & 132$\pm$26 & 210 $\pm$ 20 & 206 $\pm$ 20 & 166$\pm$6 & 175$\pm$7 \\
\textbf{SPT0441-46} & 349 & 112$\pm$22 & 118.7 $\pm$ 12 & 129.3 $\pm$ 9 & 98$\pm$6 & 106$\pm$7 \\
\textbf{SPT0459-58} & 373 & 71$\pm$14 & 81.5 $\pm$ 8 & 88.9 $\pm$ 9 & 65$\pm$6 & 80$\pm$7 \\
\textbf{SPT0459-59} & 370 & 75$\pm$15 & 77.1 $\pm$ 14 & 93.7 $\pm$ 14 & 67$\pm$7 & 75$\pm$8 \\
\textbf{SPT2132-58} & 368 & 87$\pm$17 & 81.3 $\pm$ 16 & 90 $\pm$ 13 & 75$\pm$7 & 80$\pm$7 \\
\textbf{SPT2146-55} & 354 & 117$\pm$23 & 80 $\pm$ 9 & 84.1 $\pm$ 12 & 69$\pm$12 & 83$\pm$9 \\
\textbf{SPT2311-54} & 337 & 48$\pm$10 & 120.3 $\pm$ 22 & 93 $\pm$ 16 & 106$\pm$7 & 95$\pm$7 \\
\textbf{SPT2351-57} & 436 & 36$\pm$7 & 68.4 $\pm$ 9 & 82.5 $\pm$ 8 & 56$\pm$6 & 74$\pm$6 \\
\end{tabular}
\tablefoot{$^a$ Including calibration uncertainty.
$^b$ Point Source Catalogue, see text.}
\label{table:alma-spire-comp}
\normalsize
\end{table}

\onecolumn
\begin{landscape}
\begin{longtable}{r l l l c c c c c c c}
\caption{Flux measurements.}
\label{table:measurements} \\

\toprule
Source & Label & RA & Dec & Width$\times$Height@Angle &
Flux(l+c)$^{b}$ & Flux(l)$^{b}$ & L & EW & FWHM \\
\midrule
& & & & & mJy & Jy $\cdot$ km/s & $10^8\,L_\odot$ & km/s & km/s \\
\midrule
\endfirsthead

\multicolumn{11}{c}{\tablename\ \thetable\ -- continued from previous page} \\
\toprule
Source & Label & RA & Dec & Width$\times$Height@Angle &
Flux(l+c)$^{b}$ & Flux(l)$^{b}$ & L & EW & FWHM \\
\midrule
& & & & & mJy & Jy $\cdot$ km/s & $10^8\,L_\odot$ & km/s & km/s \\
\midrule
\endhead

\midrule
\multicolumn{11}{r}{Continued on next page} \\
\endfoot

\bottomrule
\endlastfoot

SPT0106-64 & Full & 1:06:24.076 & -64:12:49.49 & $4\farcs494\times5\farcs584@90$ & $229.4 \pm 0.8$ & $<1.2$ & $<22.8$ & $<6471.4$ &  \\
SPT0106-64 & [CII]158  &             &              &                          &                 & $123.4 \pm 23.3$ & $901.5 \pm 170.1$ & & $1052 \pm 229$ \\
SPT0106-64 & A         & 1:06:24.214 & -64:12:50.11 & $1\farcs476\times0\farcs808@174$ & $63.6 \pm 3.7$  & $>-5.7^a$ & $>-103.5^a$ & $-251.8 \pm 81.2$ & $265 \pm 121$ \\
SPT0106-64 & A-small   & 1:06:24.210 & -64:12:50.13 & $0\farcs36\times0\farcs72@265$   & $35.1 \pm 8.0$  & $-3.9 \pm 1.3^a$ & $-72.0 \pm 24.2^a$ & $-121.2 \pm 35.7$ & $104 \pm 41$ \\
SPT0106-64 & B         & 1:06:23.770 & -64:12:50.03 & $1\farcs003\times1\farcs424@265$ & $123.5 \pm 3.4$ & $<5.2$ & $<95.0$ & & \\
SPT0106-64 & C         & 1:06:24.227 & -64:12:48.16 & $0\farcs36\times0\farcs64@265$   & $4.1 \pm 8.0$   & $<12.1$ & $<221.1$ & & \\
SPT0106-64 & D         & 1:06:23.920 & -64:12:50.01 & $0\farcs36\times0\farcs64@265$   & $2.9 \pm 8.0$   & $<12.1$ & $<221.1$ & $>510.1$ & $489 \pm 208$ \\
SPT0106-64 & E         & 1:06:24.144 & -64:12:48.69 & $0\farcs36\times0\farcs72@265$   & $-0.3 \pm 8.0$  & $<12.1$ & $<221.1$ & $>356.0$ & $49 \pm 20$ \\
SPT0106-64 & F         & 1:06:24.313 & -64:12:48.09 & $0\farcs36\times0\farcs72@265$   & $-1.4 \pm 8.0$  & $<12.1$ & $<221.1$ & $>565.0$ & $144 \pm 64$ \\
SPT0106-64 & G         & 1:06:24.233 & -64:12:51.44 & $4\farcs17\times4\farcs17@0$                      & $48.8 \pm 1.2$  & $<1.8$ & $<33.5$ & & \\
\hline
SPT0136-63 & Full     & 1:36:50.301 & -63:07:26.54 & $6\farcs083\times6\farcs015@102\degr$ & $51.38 \pm 0.13$ & $<0.4$  & $<6.6$    &  &  \\
SPT0136-63 & [CII]158 &             &              &                                        &                  & $29.5 \pm 1.4$ & $175.6 \pm 8.2$ &  & $482 \pm 26$ \\
SPT0136-63 & [NII]205 &             &              &                                        &                  & $0.8 \pm 0.6$  & $3.6 \pm 2.8$   &  & $198 \pm 169$ \\
SPT0136-63 & A        & 1:36:50.367 & -63:07:27.49 & $0\farcs46\times0\farcs34@161\degr$   & $4.37 \pm 2.00$ & $<6.8$  & $<100.4$  & $721.7 \pm 288.6$ & $120 \pm 51$ \\
SPT0136-63 & Arc      & 1:36:50.487 & -63:07:28.50 &                                      & $47.57 \pm 0.44$ & $<1.5$  & $<21.8$   &  &  \\
SPT0136-63 & B        & 1:36:50.167 & -63:07:21.55 & $0\farcs46\times0\farcs34@161\degr$   & $0.02 \pm 2.00$ & $2.1 \pm 0.6$ & $30.7 \pm 9.0$ & $>267.9$ & $57 \pm 24$ \\
\hline
SP0T155-62  & Full     & 1:55:48.109 & -62:50:49.88 & $2\farcs563\times2\farcs264@101\degr$ & $50.26 \pm 0.44$ & $<6.7$  & $<101.4$  &  &  \\
SPT0155-62  & [CII]158 &             &              &                                        &                  & $64.2 \pm 11.2$ & $389.3 \pm 67.8$ &  & $880 \pm 177$ \\
SPT0155-62  & [NII]205 &             &              &                                        &                  & $12.8 \pm 0.9$ & $59.6 \pm 4.0$   &  & $853.0 \pm 67$ \\
SPT0155-62  & A        & 1:55:48.043 & -62:50:50.47 & $0\farcs283\times0\farcs244@109\degr$ & $5.17 \pm 4.00$  & $<60.3$ & $<913.5$  &  &  \\
SP0T155-62  & B        & 1:55:48.214 & -62:50:49.96 &                                     & $11.83 \pm 1.69$ & $35.6 \pm 11.7$ & $539.7 \pm 177.7$ & $>237.9$ & $211 \pm 82$ \\
\hline
SPT0243-49 & Full     & 2:43:08.792 & -49:15:34.88 & $0\farcs75\times0\farcs75$           & $55.06 \pm 0.15$ & $<0.2$  & $<5.5$   & $<1963.8$ &  \\
SPT0243-49 & [CII]158 &             &              &                                        &                  & $16.9 \pm 1.6$ & $154.4 \pm 14.4$ &  & $606 \pm 66$ \\
SPT0243-49 & [NII]205 &             &              &                                        &                  & $1.5 \pm 0.5$  & $10.5 \pm 3.6$   &  & $301 \pm 120$ \\
SPT0243-49 & A        & 2:43:08.825 & -49:15:34.81 & $0\farcs261\times0\farcs208@170\degr$ & $4.87 \pm 1.00$  & $<1.5$  & $<33.1$  &  &  \\
SPT0243-49 & B        & 2:43:08.780 & -49:15:35.18 & $0\farcs261\times0\farcs208@170\degr$ & $2.62 \pm 1.00$  & $<1.5$  & $<33.1$  & $560.6 \pm 302.2$ & $314 \pm 153$ \\
SPT0243-49 & C        & 2:43:08.813 & -49:15:35.12 & $0\farcs261\times0\farcs208@170\degr$ & $1.42 \pm 1.00$  & $<1.5$  & $<33.1$  &  & $69 \pm 60$ \\
SPT0243-49 & D        & 2:43:08.787 & -49:15:35.06 & $0\farcs802\times0\farcs687@69\degr$  & $18.04 \pm 0.31$ & $<0.5$  & $<10.9$  & $179.5 \pm 50.4$ & $96 \pm 48$ \\
\hline
SPT0418-47 & Full     & 4:18:39.676 & -47:51:52.70 & $1\farcs6\times1\farcs6$            & $131.88 \pm 0.19$ & $<1.000$  & $<14.488$   & $<5994.0$ &  \\
SPT0418-47 & [CII]158 &             &              &                                      &                   & $126.8 \pm 10.4$ & $735.7 \pm 60.6$ &  & $342 \pm 32$ \\
SPT0418-47 & [NII]122 &             &              &                                      &                   & $7.2 \pm 1.8$    & $54.0 \pm 13.6$  &  & $121 \pm 35$ \\
SPT0418-47 & [NII]205 &             &              &                                      &                   & $7.9 \pm 0.7$    & $35.2 \pm 2.9$   &  & $394 \pm 38$ \\
SPT0418-47 & A        & 4:18:39.770 & -47:51:51.40 &                                 & $123.72 \pm 0.75$ & $<1.239$  & $<17.938$   &  &  \\
SPT0418-47 & B        & 4:18:39.850 & -47:51:53.54 & $0\farcs349\times0\farcs264@163\degr$ & $-0.92 \pm 2.00$  & $<10.320$ & $<149.465$  &  &  \\
SPT0418-47 & C        & 4:18:39.620 & -47:51:52.46 & $0\farcs349\times0\farcs264@163\degr$ & $0.15 \pm 2.00$   & $<10.320$ & $<149.465$  &  &  \\
SPT0418-47 & D        & 4:18:39.577 & -47:51:53.17 & $0\farcs952\times0\farcs514@59\degr$  & $7.29 \pm 0.88$   & $9.0 \pm 1.7$ & $130.0 \pm 25.1$ & $>813.5$ & $55 \pm 12$ \\
SPT0418-47 & E        & 4:18:39.581 & -47:51:51.97 & $0\farcs604\times1\farcs381@225\degr$ & $25.21 \pm 0.67$  & $<3.515$  & $<50.911$   &  &  \\
SPT0418-47 & F        & 4:18:39.668 & -47:51:53.87 & $0\farcs579\times1\farcs227@95\degr$  & $16.22 \pm 0.73$  & $<3.786$  & $<54.839$   &  &  \\
\hline
SPT0441-46 & Full     & 4:41:44.087 & -46:05:25.02 & $1\farcs0\times1\farcs0$             & $112.02 \pm 0.09$ & $<0.5$       & $<8.5$      & $<2624.0$ &  \\
SPT0441-46 & [CII]158 &             &              &                                       &                   & $22.6 \pm 2.4$ & $143.4 \pm 15.3$ &  & $471 \pm 58$ \\
SPT0441-46 & [NII]205 &             &              &                                       &                   & $1.8 \pm 0.4$  & $8.7 \pm 2.1$    &  & $351 \pm 99$ \\
SPT0441-46 & A        & 4:41:44.066 & -46:05:25.68 & $0\farcs191\times0\farcs183@15\degr$  & $9.90 \pm 1.00$  & $>-6.0^a$   & $>-94.5^a$ & $-335.3 \pm 97.6$ & $456 \pm 256$ \\
SPT0441-46 & B        & 4:41:44.150 & -46:05:25.30 &                                   & $95.42 \pm 0.17$ & $<1.0$      & $<16.1$   &  &  \\
SPT0441-46 & C        & 4:41:44.076 & -46:05:24.33 & $0\farcs393\times0\farcs662@254\degr$ & $14.74 \pm 0.36$ & $<2.1$      & $<33.8$   &  &  \\
SPT0441-46 & D        & 4:41:44.142 & -46:05:25.99 & $0\farcs191\times0\farcs183@15\degr$  & $-0.84 \pm 1.00$ & $4.8 \pm 1.2$ & $75.5 \pm 19.7$ & $>128.1$ & $432 \pm 139$ \\
\hline
SPT0459-58 & Full     & 4:58:59.782 & -58:05:14.23 & $1\farcs0\times1\farcs0$             & $70.55 \pm 0.98$ & $25.0 \pm 7.9$  & $448.3 \pm 141.2$ & $478.1 \pm 145.5$ & $448 \pm 171$ \\
SPT0459-58 & [CII]158 &             &              &                                       &                  & $25.3 \pm 1.8$  & $182.0 \pm 13.0$  &                   & $345 \pm 28$ \\
SPT0459-58 & [NII]205 &             &              &                                       &                  & $2.6 \pm 0.8$   & $14.1 \pm 4.2$    &                   & $401 \pm 143$ \\
SPT0459-58 & A        & 4:58:59.843 & -58:05:14.41 & $0\farcs372\times0\farcs296@146\degr$ & $4.68 \pm 6.00$  & $>-3.7^a$      & $>-65.5^a$       & $-343.9 \pm 103.9$ & $149 \pm 110$ \\
SPT0459-58 & B        & 4:58:59.776 & -58:05:14.00 & $0\farcs372\times0\farcs296@146\degr$ & $5.17 \pm 6.00$  & $<3.7$         & $<65.5$          &                   &  \\
\hline
SPT0459-59 & Full     & 4:59:12.245 & -59:42:20.72 & $2\farcs0\times2\farcs0$           & $75.10 \pm 0.17$ & $<0.5$         & $<8.4$           &                   &  \\
SPT0459-59 & [CII]158 &             &              &                                      &                  & $57.7 \pm 2.9$ & $407.2 \pm 20.5$ &                   & $558 \pm 33$ \\
SPT0459-59 & [NII]205 &             &              &                                      &                  & $4.1 \pm 0.7$  & $22.0 \pm 3.9$   &                   & $446 \pm 92$ \\
SPT0459-59 & A        & 4:59:12.136 & -59:42:21.70 & $0\farcs38\times0\farcs297@143\degr$ & $2.20 \pm 2.00$ & $2.1 \pm 0.7$  & $36.2 \pm 11.5$  & $1048.5 \pm 357.0$ & $118 \pm 44$ \\
SPT0459-59 & B        & 4:59:12.294 & -59:42:19.80 & $0\farcs38\times0\farcs297@143\degr$ & $3.00 \pm 2.00$ & $>-5.4^a$      & $>-95.1^a$       & $-367.0 \pm 132.4$ & $90 \pm 52$ \\
\hline
SPT2132-58 & Full      & 21:32:43.231 & -58:02:46.21 & $0\farcs7\times0\farcs7$             & $87.34 \pm 0.07$ & $<0.1$        & $<1.3$          &                   &  \\
SPT2132-58 & [CII]158  &               &               &                                       &                  & $38.0 \pm 8.0$ & $265.7 \pm 55.9$ &                   & $212 \pm 52$ \\
SPT2132-58 & A         & 21:32:43.244 & -58:02:46.41 & $0\farcs373\times0\farcs193@140\degr$ & $21.74 \pm 0.37$ & $<0.4$        & $<6.8$          &                   &  \\
SPT2132-58 & A-small   & 21:32:43.246 & -58:02:46.40 & $0\farcs109\times0\farcs083@131\degr$ & $4.39 \pm 1.00$  & $<1.1$        & $<18.5$         & $230.0 \pm 58.2$ & $148 \pm 78$ \\
SPT2132-58 & B         & 21:32:43.283 & -58:02:46.06 & $0\farcs109\times0\farcs083@131\degr$ & $0.34 \pm 1.00$  & $<1.1$        & $<18.5$         & $>481.6$         & $130 \pm 56$ \\
SPT2132-58 & C         & 21:32:43.238 & -58:02:45.82 & $0\farcs109\times0\farcs083@131\degr$ & $1.84 \pm 1.00$  & $<1.1$        & $<18.5$         & $643.6 \pm 225.0$ & $163 \pm 70$ \\
\hline
SPT2146-55 & Full     & 21:46:54.048 & -55:07:54.43 & $1\farcs4\times1\farcs4$               & $117.03 \pm 0.0066$ & $<0.04$       & $<0.7$          & $<176.1$ &  \\
SPT2146-55 & [CII]158 &               &               &                                         &                     & $44.0 \pm 9.0$ & $287.9 \pm 59.2$ &          & $302 \pm 72$ \\
SPT2146-55 & [NII]205 &               &               &                                         &                     & $5.0 \pm 0.5$ & $25.3 \pm 2.5$  &          & $413 \pm 48$ \\
SPT2146-55 & A        & 21:46:53.962 & -55:07:54.05 & $0\farcs424\times0\farcs204@133\degr$  & $14.16 \pm 0.064$  & $6.2 \pm 2.1$ & $101.4 \pm 34.5$ & $334.8 \pm 78.4$ & $160 \pm 64$ \\
SPT2146-55 & B        & 21:46:53.975 & -55:07:53.94 & $0\farcs101\times0\farcs088@112\degr$  & $0.32 \pm 0.2$     & $1.2 \pm 0.3$ & $20.0 \pm 5.6$  & $>1304.2$ & $142 \pm 47$ \\
SPT2146-55 & C        & 21:46:53.956 & -55:07:54.11 & $0\farcs101\times0\farcs088@112\degr$  & $4.12 \pm 0.2$     & $<1.2$        & $<19.6$         &          &  \\
SPT2146-55 & D        & 21:46:54.137 & -55:07:54.76 & $0\farcs311\times0\farcs185@120\degr$  & $15.11 \pm 0.25$   & $<0.5$        & $<7.7$          &          &  \\
\hline
SPT2311-54 & Full     & 23:11:23.960 & -54:50:30.38 & $0\farcs8\times0\farcs8$ & $48.48 \pm 0.31$ & $<0.2$         & $<3.1$           & $<1951.5$          &               \\
SPT2311-54 & [CII]158 &               &               &                         &                  & $51.2 \pm 5.4$ & $303.0 \pm 32.0$ &                     & $360 \pm 44$ \\
SPT2311-54 & [NII]205 &               &               &                         &                  & $1.9 \pm 1.2$  & $8.6 \pm 5.4$   &                     & $472 \pm 342$ \\
SPT2311-54 & A        & 23:11:23.947 & -54:50:30.82 & $0\farcs835\times0\farcs659@166\degr$ & $9.37 \pm 0.68$ & $<0.4$          & $<6.6$            & $1353.6 \pm 510.8$ & $389 \pm 183$ \\
SPT2311-54 & B        & 23:11:23.967 & -54:50:30.18 & $0\farcs608\times0\farcs508@54\degr$  & $22.36 \pm 0.90$ & $<0.4$          & $<6.6$            &                     &               \\
\hline
SPT2351-57 & Full     & 23:51:50.798 & -57:22:18.48 & $2\farcs252\times1\farcs783@37\degr$ & $35.84 \pm 0.37$ & $<0.5$        & $<12.1$          & $<9922.7$          &               \\
SPT2351-57 & [CII]158 &               &               &                                         &                  & $22.2 \pm 4.4$ & $209.1 \pm 41.5$ &                     & $660 \pm 151$ \\
SPT2351-57 & [NII]205 &               &               &                                         &                  & $<0.03$        & $<0.7$           &                     &               \\
SPT2351-57 & A        & 23:51:50.809 & -57:22:18.70 & $0\farcs685\times0\farcs493@180\degr$ & $4.26 \pm 1.28$ & $1.6 \pm 0.4$  & $37.5 \pm 10.0$  & $462.8 \pm 123.9$  & $225 \pm 71$ \\
SPT2351-57 & B        & 23:51:50.739 & -57:22:18.36 & $0\farcs404\times1\farcs32@250\degr$  & $15.06 \pm 1.01$ & $>-1.4^a$      & $>-32.9^a$       & $-100.5 \pm 43.5$  & $196 \pm 104$ \\
SPT2351-57 & C        & 23:51:50.777 & -57:22:18.23 & $0\farcs6\times0\farcs326@171\degr$   & $10.16 \pm 1.70$ & $>-2.3^a$      & $>-54.1^a$       & $-74.1 \pm 29.4$   & $214 \pm 125$ \\
SPT2351-57 & D        & 23:51:50.706 & -57:22:18.48 & $0\farcs6\times0\farcs326@171\degr$   & $1.25 \pm 1.70$  & $>-2.3^a$      & $>-54.1^a$       & $-404.5 \pm 159.7$ & $83 \pm 60$ \\

\end{longtable}
\tablefoot{$^{a}$ Lower limit due to negative continuum-subtracted line flux, possibly indicating absorption. $^{b}$ Excluding flux uncertainty.}

\end{landscape}
\twocolumn

\section{Notes on individual sources}\label{section:individual-sources}
We here provide an overview of each source, explaining our motivation for defining the apertures listed in Table\,\ref{table:measurements}.

\subsection{SPT0106-64}
This source consists of an incomplete lensing arc (aperture G), surrounded by two very bright continuum sources (apertures A and B). Within the arc, we add an aperture C on the continuum peak, as well as three apertures D, E, and F where some tentative narrow \OI\ emission may escape.

\subsection{SPT0136-63}
This source consists of a curved arc with several continuum peaks. Aperture A marks a small gap in the continuum emission where some tentative \OI\ emission escapes. Apertures B and C show even more tentative emission at the prolongation of the arc where also some Band 7 emission was detected (B).

\subsection{SPT0155-62}
This source was detected at the very edge of the Band 10 primary beam, leading to a strongly varying sensitivity over the Einstein ring, and less reliable photometry. We detect no \OI\ at the continuum peak aperture A, but tentative emission in aperture B. Deeper data at the phase centre are required to confirm this.

\subsection{SPT0243-49}
This is one of the two sources observed in Band 9, leading to slightly worse spatial resolution. No \OI\ is detected at the continuum peak aperture A. We do find tentative emission in the larger aperture D, which we investigate deeper in the smaller apertures B and C; especially in the latter, the narrow emission appears stronger.

\subsection{SPT0418-47}
This well-studied source consists of a full Einstein Ring (aperture A). We adopt the nomenclature of \citep{peng2025a} for the companion source components B and C. We also add apertures C, E and F on the continuum peaks in the Einstein ring. Intriguingly, aperture D in the main continuum gap of the ring has the strongest \OI\ emission.

\subsection{SPT0441-46}
This source consists of two bright parts of an Einstein ring. Tentative \OI\ absorption could be seen against the brightest continuum clump A, with emission in component D at the same velocity range.

\subsection{SPT0459-58}
This source is the only one showing an \OI\ detection in the Full aperture. The brightest continuum aperture A shows tentative \OI\ absorption, which is not seen in the other aperture B.

\subsection{SPT0459-59}
This source is composed of many clumps. Aperture A shows tentative \OI\ in emission, which aperture B shows it in absorption at a similar velocity offset.

\subsection{SPT2132-58}
This source consists of a bright continuum component A, showing no \OI/ in emission nor absorption, though some tentative emission may be seen in a small aperture. Other tentative emission near the well-defined \CII\ redshift is seen in apertures B and C.

\subsection{SPT2146-55}
This source was observed later in Cycle~11 with a larger antenna configuration, meaning it is potentially over-resolved. Only the brightest continuum components A and D detected in continuum. We split aperture A into apertures B and C, with the former showing some narrow \OI\ emission in B, but none in C.

\subsection{SPT2311-54}
The bright continuum component B shows no \OI\, but the more diffuse emission in aperture A shows tentative \OI\ covering almost the entire \CII\ line width.

\subsection{SPT2351-57}
This compact source is barely resolved in the continuum, but some tentative \OI\ emission is seen at the edge of the source in aperture A. We define a double aperture B showing tentative \OI\ absorption, which we split in apertures C and D.

\end{appendix}

\end{document}